\documentclass[preprint]{aastex}

\shorttitle{Radiation on Disks}
\shortauthors{F. C. Adams}

\newcommand{\be}{\begin{equation}}
\newcommand{\ee}{\end{equation}}
\newcommand{\muc}{ \langle \mu \rangle}

\newcommand{\sigstar}{{\sigma_{FUV} }}

\newcommand{\heat}{ {\cal H} } 
\newcommand{\nstar}{ {N_\star} } 
\newcommand{\tevap}{ {t_{\rm evap}} } 
\newcommand{\atau}{ \tau_{FUV} } 
\newcommand{\atauess}{ \tau_{s\infty} } 

\def\lta{\,\raise 0.3 ex\hbox{$ < $}\kern -0.75 em
 \lower 0.7 ex\hbox{$\sim$}\,}
\def\gta{\,\raise 0.3 ex\hbox{$ > $}\kern -0.75 em
 \lower 0.7 ex\hbox{$\sim$}\,} 

\begin{document}

\title{PHOTOEVAPORATION OF CIRCUMSTELLAR DISKS \\
DUE TO EXTERNAL FUV RADIATION IN STELLAR AGGREGATES}

\author{Fred C. Adams$^{1,2}$,
David Hollenbach$^3$, Gregory Laughlin$^4$, Uma Gorti$^3$}

\affil{$^1$Michigan Center for Theoretical Physics, Physics Department \\ 
University of Michigan, Ann Arbor, MI 48109}

\affil{$^2$Astronomy Department, University of Michigan, Ann Arbor, MI 48109}

\affil{$^3$NASA Ames Research Center, Moffett Field, CA 94035}

\affil{$^4$Lick Observatory, University of California, Santa Cruz, CA 95064}

\begin{abstract}

When stars form within small groups (with $\nstar \approx 100 - 500$
members), their circumstellar disks are exposed to relatively little
EUV ($h\nu > 13.6$ eV) radiation but a great deal of FUV (6 eV $< h\nu
<$ 13.6 eV) radiation ($\sim10^3$ times the local interstellar FUV
field) from the most massive stars in the group. This paper calculates
the mass loss rates and evaporation time scales for circumstellar
disks exposed to external FUV radiation.  Previous work treated large
disks and/or intense radiation fields in which the disk radius $r_d$
exceeds the critical radius $r_g$ where the sound speed in the FUV
heated surface layer exceeds the escape speed; it has often been
assumed that photoevaporation occurs for $r_d > r_g$ and is negligible
for $r_d < r_g$.  Since $r_g \gta 100$ AU for FUV heating, this would
imply little mass loss from the planet-forming regions of a disk. In
this paper, we focus on systems in which photoevaporation is
suppressed because $r_d < r_g$ and show that significant mass loss
still takes place as long as $r_d/r_g \gta 0.1 - 0.2$. Some of the gas
extends beyond the disk edge (or above the disk surface) to larger
distances where the temperature is higher, the escape speed is lower,
and an outflow develops. The resulting evaporation rate is a sensitive
function of the central stellar mass and disk radius, which determine
the escape speed, and the external FUV flux, which determines the
temperature structure of the surfaces layers and outflowing gas. Disks
around red dwarfs, low mass stars with $M_\ast \lta$ 0.5 $M_\odot$,
are evaporated and shrink to disk radii $r_d \lta 15$ AU on short time
scales $t \lta 10$ Myr when exposed to moderate FUV fields with $G_0$
= 3000 (where $G_0$ = 1.7 for the local interstellar FUV field).  The
disks around solar type stars are more durable. For intense FUV
radiation fields with $G_0$ = 30,000, however, even these disks shrink
to $r_d \lta 15$ AU on time scales $t \sim10$ Myr. Such fields exist
within about 0.7 pc of the center of a cluster with $N_\star \approx
4000$ stars. If our solar system formed in the presence of such strong
FUV radiation fields, this mechanism could explain why Neptune and
Uranus in our solar system are gas poor, whereas Jupiter and Saturn
are relatively gas rich. This mechanism for photoevaporation can also
limit the production of Kuiper belt objects and can suppress giant
planet formation in sufficiently large clusters, such as the Hyades,
especially for disks associated with low mass stars.

\end{abstract}

\section{INTRODUCTION}

The collapse of molecular cloud cores leads to the formation of stars
with orbiting accretion disks.  The dust in these disks can settle,
coagulate, and form solid objects ranging in sizes from pebbles to
planetesimals to planets.  However, a number of mechanisms act to
disperse gas from these disks, either driving the gas back out into
the interstellar medium, or spiraling it into the central star.
During this dispersal, the gas can, in turn, drag small dust particles
(with radii $b \lta 1$ cm) along in the flow.  Gas dispersal thus
disrupts planet formation in at least two important ways: (i) If the
gas is dispersed before the disk dust particles have coagulated to
sizes sufficient to decouple from the gas flow ($b \gta 1$ cm), then
the formation of planetesimals, Kuiper Belt Objects, and rocky planets
will be curtailed because all the orbiting solid material in the gas
flow region is removed before it has a chance to grow.  (ii) If the
gas is dispersed before large ($m_P \gta 5-15$ earth masses) rocky
planets are formed, and if giant planets form by the gravitational
accretion of gas onto these large rocky cores (Bodenheimer \& Pollack
1986; Lissauer 1993), then the formation of gas giant planets like
Jupiter and Saturn will be suppressed.

The dispersal of gas and small dust particles has other important
effects on the formation of planetary systems. The presence of a
moderately massive gas disk leads to planetary migration (Lin \&
Papaloizou 1986, Ward 1997). The presence of even small amounts of gas
at time scales $t \sim 10 - 100$ Myr after disk formation
influences the dynamics and evolution of orbiting objects in solar
systems. For example, such gas can affect the orbital eccentricities
of both planets and any remaining planetesimals (e.g., Tanaka \& Ida
1997; Kominami \& Ida 2002; Chiang et al. 2002). Reducing the
eccentricities can, in turn, alter the time required for the
collisional agglomeration of large planets.

Observations of disk systems of various ages suggest that the small
($b \lta 1$ mm) dust particles disappear on time scales of roughly 3
-- 10 Myr (Haisch et al. 2001). Near infrared continuum observations
probe dust orbiting in the central regions of disks, $r \lta 0.1$ AU,
whereas submillimeter and millimeter wavelength continuum observations
probe dust in the outer disks, $r \gta 30$ AU.  The small dust grains
in both regions disappear on roughly the same time scale. Presumably,
after the disappearance time scale, some of the small dust particles
have coagulated to form pebbles and larger objects that are no longer
detectable at IR or millimeter wavelengths.  Several authors cite
observational evidence for coagulated dust of size $b \sim 1$ cm in
young disks (d'Alessio et al. 1999, Throop et al. 2001).  However, the
present observations do not specify how much of the small dust has
been dispersed, and how much has coagulated into larger objects and
disappeared from view.

Observations of gas in disks indicate that gas can also be dispersed
in a relatively short time of only $t \lta 10$ Myr (Zuckerman et
al. 1995). Generally, the gas is traced by millimeter and
submillimeter observations of the trace species $^{12}$CO in the
J=1-0, 2-1, or 3-2 transitions, which are sensitive to low masses of
gas as long as the gas disk is extended.  Because these lines are
optically thick, and beam dilution reduces the observed intensity for
these small ($\lta 1"$) disks, the current surveys are insensitive to
gas of any mass at $r \lta 50$ AU for the nearby ($d \sim 100$ pc)
young star/disk systems.  Nevertheless, significant gas masses have
been detected via CO observations in disks as old as 10 Myr
(cf. Carpenter 2002). In short, from an observational point of view,
it appears that the bulk of the gas is dispersed from the outer disks
in time scales $t \lta 10$ Myr, but the evolution of the gas in the
inner, planet-forming region of the disk is uncertain.

Hollenbach, Yorke, \& Johnstone (2000) reviewed theoretical models for
dispersing the gas and small dust from disks.  Observational evidence
from our solar system and other planetary systems indicates that more
gas and dust is accreted onto the central star and dispersed back into
the ISM than forms planets or other solid orbiting objects. This
dispersal is dominated by photoevaporation in the outer regions of
disks and viscous evolution (accretion onto the star coupled with
protostellar outflows) in the inner parts of disks. The boundary
between these two regimes -- viscous evolution and photoevaporation --
remains uncertain.  We need to develop a better understanding of
viscous accretion and develop better photoevaporation models that
accurately track mass loss at moderate radii from the central star.
This paper addresses the latter problem for the case of external
irradiation.  Stellar winds may play a significant, but probably not
dominant, role in dispersing gas at moderate radii near the boundary
of the inner viscosity-dominated region and the outer photoevaporating
region.  Nearby stellar encounters, even for star/disks born in dense
clusters like the Trapezium cluster, only affect the outermost regions
($r\gta 100$ AU) of the largest disks, and, even there, the
photoevaporation of disks in these same clusters is likely to dominate
the dispersal of the outer regions (e.g., Scally \& Clarke 2001,
Clarke 2002, Adams \& Laughlin 2001).

Photoevaporation occurs when energetic photons heat the surface of the
disk to elevated temperatures. The radiation of interest includes FUV
photons in the energy range 6eV -- 13.6eV, EUV photons in the energy
range 13.6eV -- 100eV, and X-rays in the energy range 100eV -- 10keV.
If EUV photons can penetrate the outward flow and reach the disk
surface, they will ionize and heat the surface to $T \approx 10^4$ K,
whereas the FUV and/or X-ray photons tend to heat the neutral gas to
lower temperatures, typically in the range 100 K $< T < $3000 K. The
thermal pressures in these heated regions drive the gas outward and
create a flow into the interstellar medium. An important critical
radius $r_g$ can be defined -- this fiducial length scale is the
radius where the sound speed of the gas (hydrogen atoms) equals the
escape speed from the gravitationally bound system, i.e.,  
\be
r_g = {G M_\ast \muc \over k T} \approx 100 {\rm AU}
\, \Bigl( {T \over 1000 {\rm K} } \Bigr)^{-1}
\, \Bigl( {M_\ast \over 1 M_\odot } \Bigr) \, ,
\label{eq:rcrit} 
\ee
where $M_\ast$ is the mass of the central star and $\muc$ is the
average mass of the gas particles. Most previous work on
photoevaporation (Hollenbach et al. 1994, Johnstone et al.  1998,
St\"orzer \& Hollenbach 1999) assumed that photoevaporation flow is
only active for $r > r_g$, and that the disk is static (with a warm
surface corona held in orbit by the stellar gravity) for $r < r_g$. 
In this paper, we generalize this picture to include a more proper
treatment of the flow hydrodynamics and show that significant
photoevaporation can take place for smaller radii, $r \gta 0.2 r_g$
(see Figure 1). In any event, disk photoevaporation can be considered
like a slow ($v \sim 1-5$ km s$^{-1}$) thermal Parker wind originating 
from the outer portion of the disk ($r \sim 3 - 100$ AU). 

Other authors have discussed the possibility of significant flow from
$r<r_g$ for disks surrounding compact objects (e.g., Begelman, McKee
\& Shields 1983; Woods et al. 1996). This work showed that
significant photoevaporation can take place outwards from $r \sim 0.2
r_g$ for the case of X-ray heated disks around black holes. Recently,
Liffman (2003) presented an analytic argument for photoevaporative
flow inside of $r_g$. Flow inside $r_g$ is important for protoplanetary 
disks.  Often the heating raises the disk surfaces to $T \lta 1000$ K,
so that $r_g \gta 100$ AU. If photoevaporative flows are still
significant at $r \sim 0.2 r_g$, then photoevaporation can effectively
remove gas and dust from the region near 20 AU and thereby affect the
formation of Uranus, Neptune, and Kuiper Belt Objects in our solar
system.

Photoevaporation can be initiated by the energetic photons from the
central star or from a nearby, more massive, and luminous star in the
stellar birth cluster.  Hollenbach et al. (1994) originally modeled
the evaporation caused by EUV photons from the central star.
Johnstone, Hollenbach \& Bally (1998) and St\"orzer \& Hollenbach
(1999) presented the first semi-analytic models of disks (with $r_d >
r_g$) around low mass stars being photoevaporated by the FUV and EUV
fluxes from a nearby OB star.  These models were successfully applied
to the PROtoPLanetarY DiskS, or ``proplyds'', observed in the cluster
of low mass stars around the Trapezium in Orion (e.g., O'Dell 1998,
Bally et al. 1998, Churchwell et al. 1987). A complementary set of
models (Richling \& Yorke 1997, 1998, 2000; Yorke \& Richling 2002)
studied the hydrodynamical flow for disks subjected to both radiation
from their central stars and external radiation.  This previous work
produced two results of interest here: (i) In the case of external
illumination, the FUV photons often initiate the mass loss and the
incident EUV flux is absorbed at an ionization front in the neutral
flow which is several disk radii away from the disk surface. (ii) The
externally-illuminated disks evaporate from outside in, whereas the
bulk of the mass loss for internally-EUV-illuminated disks occurs at
$r \sim r_g$. In other words, in the former (external) case, a disk
with outer radius $r_d$ shrinks from $r_d > r_g$ to $r_d \lta r_g$ as
evaporation proceeds.  In the latter case, in the absence of turbulent
viscosity to drive radial flow and replenish material at $r_g$, the
disk evaporates at $r_g$ until a gap is formed there, and then the
photoevaporation proceeds from $r_g$ outward to $r_d$.  These early
models effectively assumed large disks with $r_d > r_g$.  This paper
presents a more in depth treatment for the case of small disks (with
$r_d < r_g$) that are externally illuminated by FUV radiation.
However, this work also has important implications for
photoevaporation at $r<r_g$ for the internally-illuminated disks.

Several recent papers combine these early photoevaporation models with
models of viscous accretion in attempts to model the time evolution of
the dispersal of the entire protoplanetary disk. Clarke et al. (2001)
treat the EUV photoevaporation by the central star coupled with
viscous accretion and evolution to explain why disks are observed to
rapidly disperse at the end of their lives on a time scale that is a
small fraction of the disk lifetime. Matsuyama et al. (2003ab) model
disks with both internal EUV and external FUV and EUV illumination and
with viscous evolution.  Such models will need modification in light
of the results of this paper.

Stars often form in groups or clusters. If these stellar aggregates
are large enough ($\nstar \gta 200$ stars), the system has a good
chance of containing at least one O or early B star. In such systems,
the low mass stars in the cluster are subject to significant
photoevaporation by the FUV flux from other, larger stellar members.
The ultimate goal for the work presented in this paper is to calculate
the probability that a given low-mass star/disk system in the Galaxy,
like the early solar nebula, experienced sufficient external
illumination and consequent photoevaporation to affect the formation
of gas-giant planets and/or the formation of planetesimals and planets
in the Kuiper Belt region.  Such a calculation would require the
knowledge of the probability of being born in a cluster of given size
$\nstar$ (e.g., see Lada \& Lada 2003, Porras et al. 2003, Adams \&
Myers 2001, Carpenter 2000), the probability that one or several high
mass stars are members of this cluster (given by the stellar initial
mass function), the delay time between low mass star formation and
high mass star formation, and the fraction of time that the low mass
stars lie at a given distance from the unembedded OB star. Armitage
(2000) presents a first attempt at such a model, which assumes that
EUV-induced photoevaporation operates only for $r_d > r_g$.  This
paper presents the corresponding analysis for the case of FUV-induced
photoevaporation that occurs for $r_d < r_g$; this latter process
often dominates the mass loss for typical disks in typical star
formation environments.
 
This paper is organized as follows. In \S 2, we discuss the physical
mechanisms in photoevaporating disks, including heating processes,
dust properties and attenuation, cooling mechanisms, thermal balance,
and chemistry. We then summarize in more detail (in \S 3) the previous
results for the photoevaporation of ``large'' (supercritical) disks
with $r_d > r_g$, since these analytic results will be useful for
generalization to the case of smaller disks. In \S 4, we calculate the
photoevaporative mass loss rates and time scales for subcritical disks
(with $r_d < r_g$) due to external FUV illumination. In general,
photoevaporation takes place on both the disk surface, creating an
initially vertical flow, and from the disk edge at $r_d$, creating a
radial flow.  Although the disk edge has less area, the radial flow
tends to dominate the mass loss because the material here is bound
more weakly.  In a previous paper (Hollenbach \& Adams 2003), we
presented the isothermal case, where we can obtain analytic
approximations which provide physical insight; here we develop the
more complicated (but more realistic) non-isothermal case where the
temperature is determined from the heating and cooling of the gas in
the flow. We determine how the mass loss rate depends on the incident
FUV flux, the size $r_d$ of the disk, and the mass of the central
star.  We apply these results to the possible evaporation of the early
solar nebula (\S 5), the formation of Kuiper Belt objects and debris
dust (\S 6), the suppression of giant planet formation in large
clusters like the Hyades (\S 7), and the evaporation of disks around
low mass stars (\S 7).  We conclude, in \S 8, with a summary and
discussion of our results.

\section{PHYSICAL MECHANISMS IN PHOTOEVAPORATING DISKS}

\subsection{Overview} 

In an ideal case, one would solve the photoevaporation problem using a
full three-dimensional treatment of the hydrodynamics, including
time-dependent heating, cooling, and chemistry. Unfortunately,
however, such a calculation is beyond the scope of this initial
effort. Instead, we numerically solve the streamline equation for the
flow hydrodynamics (in the spherical approximation) by utilizing
temperatures derived from a state-of-the-art photodissociation region
(PDR) code (Kaufman et al. 1999).  This code self-consistently solves
for the chemical abundances and gas temperature at any position
(defined by the column density of hydrogen $N_H$ from that position to
the FUV source) and for a given hydrogen gas density $n$. The PDR code
assumes that thermal balance (heating and cooling rates are equal) and
steady state chemical abundances have been achieved.  We can check,
{\it post facto}, that the flow time scales are long enough to justify
these approximations. As we discuss below, when the flow approaches
the sonic point, the ratio of the flow time to the heating time
becomes smaller. We assume here that the gas temperature approaches a
constant value near the sonic point and in the outer region of the
flow.

In order to outline the physical mechanisms that operate during
photoevaporation, we must define the basic flow quantities and their
benchmark values. The outer radius of the disk $r_d$ marks the inner
boundary of the flow (and the inner boundary of our calculations); we
are primarily interested in disks of roughly solar system size, with
$r_d$ = 20 -- 60 AU, and with moderate FUV heating so that $r_d <
r_g$.  The flow begins subsonically, with $v \ll a_S$, at the inner
boundary $r_d$. As the gas flows outwards, the temperature and hence
the sound speed increases. The flow speed increases more rapidly so
that the the Mach number $\cal M$ = $v/a_S$ increases and the flow
reaches a sonic point. The radius $r_s$ of the sonic point is
typically comparable to (but smaller than) the fiducial radius $r_g$
defined above (eq. [\ref{eq:rcrit}]). For typical cases considered
here (see \S 4), the sonic radius is a few times the outer disk edge
and the critical radius is a few times the sonic radius.  This paper
thus works in the (previously unstudied) regime where $r_d/r_g
\sim 0.2$ and finds that substantially mass loss can still take place 
for moderate values of the external FUV radiation field. 

The results of this paper are primarily applicable to cases where a
low mass star/disk system is formed within a small cluster or stellar
group with $\nstar$ = 100 -- 500 stars (see, e.g., Lada \& Lada 2003,
Porras et al. 2003, Adams \& Myers 2001). In this setting, the disk
will typically be illuminated by an O or B star which lies within the
stellar birth aggregate, at a distance of 0.1 -- 1 pc (e.g., Testi et
al. 1997, 1998, 1999). For example, at a distance of 0.3 pc, a single
main-sequence star of mass $M_\ast$ = 8.7 $M_\odot$ will produce an
FUV radiation field with $G_0 \approx 2000$ (Parravano, Hollenbach, \&
McKee 2003).\footnote{Throughout this paper, we follow the standard
convention of using the dimensionless parameter $G_0$ to measure the
incident FUV flux. Specifically, $G_0=1$ corresponds to a radiative
flux 1.6$\times 10^{-3}$ erg cm$^{-2}$ s$^{-1}$ in the 912\AA \ to
2000\AA \ band; this benchmark flux is typical of the local
interstellar radiation field.} According to the standard initial mass
function, a stellar aggregate with $\nstar \approx 200$ will have a
50-50 chance of producing a star this large (Adams \& Myers 2001).
Notice that the total FUV radiation field produced by all the stars in
the aggregate will be somewhat larger, so that $G_0 \approx 3000$ is a
reasonable benchmark value. We also stress that because stellar
aggregates of this size will have relatively few massive stars, the
radiation fields they produce will vary substantially from group to
group due to incomplete sampling of the IMF.

These results also apply to low mass star/disk systems in large
clusters, since even in the presence of EUV radiation, the FUV
generally dominates the photoevaporation process until the disks
shrink to sizes $r_d \lta 10 - 20$ AU (see St\"orzer \& Hollenbach
1999 and the discussion in \S 5). For a cluster containing $N_\star
\approx 4000$ stars, similar to the Trapezium cluster, the OB stars 
will typically produce a field $G_0 \approx 13,000$ $d_{\rm pc}^{-2}$, 
where $d_{\rm pc}$ is the distance in parsecs to the cluster center
(Parravano 2003, private communication). Thus, for these large
clusters, we take $G_0 \approx 30,000$ as a benchmark value
(applicable to stars in the cluster core with $d_{\rm pc} < 1$).

Although this paper treats photoevaporation by calculating the depth
dependent temperature of the gas, it is useful to keep in mind a
simplified model of the photoevaporation mechanism: A surface layer of
gas is heated to to a fixed temperature $T$ and a flow develops until
the column density of the flow reaches a critical column density
$N_C$. In this idealized model, the two flow parameters ($T, N_C$),
along with the stellar mass $M_\ast$ and disk radius $r_d$, completely
specify the mass loss rate. As we discuss below, the column density is
approximately that required to make the flow optically thick at FUV
wavelengths so that $N_C \sim 10^{21}$ cm$^{-2}$.  The typical density
at the base of the flow is about $n_d \sim 10^7$ cm$^{-3}$, which is
larger than, but comparable to the naive estimate $n \sim N_C/r_d$.
The Appendix presents a modified version of the analytic treatment
given in Hollenbach \& Adams (2003) for the photoevaporative mass loss
rates from disks with $r_d < r_g$ (where the external heating is
approximated by a semi-analytic model).

\subsection{Gas Heating Mechanisms}

Heating of the gas on the surface or edges of the disk ultimately
causes the photoevaporative flow. The gas heating itself is driven by
incident energetic photons.  In this paper we focus on the mass loss
driven by heating from FUV photons, although we note that EUV photons
will also be present and can instigate additional mass loss (see the 
discussion in \S 5).  

The FUV heating of the neutral gas at the disk surface and in the
photoevaporative flow arises mainly from two mechanisms: grain
photoelectric heating and the FUV pumping and subsequent collisional
de-excitation of H$_2$ molecules.  In the former process, the FUV
photons are absorbed by dust grains and a small fraction of these
absorption events leads to the ejection of an energetic electron ($E
\sim$ 1 eV) into the gas.  As the electron collides with gas atoms and
ions, it shares its kinetic energy as heat. In typical cases, about
1\% of the absorbed FUV photon energy is delivered as gas heating
through this mechanism. In the latter process, H$_2$ molecules absorb
FUV photons at particular electronic transition wavelengths in the
range 912\AA $\le \lambda \le$ 1100\AA. These absorption events lead
to electronic excitation of H$_2$, followed by fluorescent decay to
bound vibrational states of the ground electronic state (90\% of the
time).  At the high densities at or near disk surfaces, this
vibrational energy is converted to heat via collisional de-excitation
by other hydrogen atoms or molecules (for further detail, see
Hollenbach \& Tielens 1999).

\subsection{Dust Properties}

Dust plays several important roles in the photoevaporation process.
The attenuation of the FUV photons by small ($b \sim 0.001-0.1 \mu$m)
dust particles carried along in the gas flow limits the depth of the
FUV heating. Dust also provides heating processes which (in part)
determine the gas temperature, either directly by grain photoelectric
heating, or indirectly by providing the chemical catalyst that forms
H$_2$ (which affects both the heating and cooling rates).

Dust can be the agent that determines this critical column density
$N_C$ (see \S 2.1 and Johnstone et al. 1998). In particular, dust
properties determine $N_C$ for high ratios of the FUV flux to the gas
density $G_0/n \gta 10^{-2}$ cm$^3$. Such high ratios ensure that
H$_2$ self-shielding is not very effective at the surface, so that the
H$_2$ abundance (and other molecular species that follow) increases
only when the dust optical depth $\tau_{FUV}$ at FUV wavelengths
becomes significant. In this case, $N_C = N_{FUV}$, the column density
required for $\tau_{FUV} = 1$, i.e., $N_{FUV} \sim 10^{21}$
cm$^{-2}$. For larger optical depths, heating by FUV photons is less
efficient and cooling rates via molecular species grow larger; as a
result, the temperature drops precipitously.

The rough criterion that the dust optical depth of the flow is of
order unity is easy to understand. If the column density $N_H$ from
the base of the flow to the FUV source were small so that $\tau_{FUV}
\ll 1$, then the FUV radiation would penetrate further and heat higher
density gas deeper in the disk. This penetration would raise the
column density $N_H$ and the corresponding optical depth $\tau_{FUV}$
of the flow.  On the other hand, if $N_H$ were large so that
$\tau_{FUV} \gg 1$, then FUV photons could not penetrate to the base
of the flow, the gas would not be heated there, and the solution would
be inconsistent because no flow could originate from such cold gas
deep inside the disk.

In this work, we consider systems with lower FUV fluxes and higher gas
densities than for cases considered previously (like the proplyds
modeled in Johnstone et al. 1998), so that $G_0/n$ can be less than
$10^{-2}$ cm$^3$. In this regime, the H$_2$ self shields and the gas
becomes predominantly H$_2$ at an optical depth less than $\tau _{FUV}
= 1$, or a column less than the benchmark value $N_{FUV}\sim 10^{21}$
cm$^{-2}$. The gas then cools {\it before} the FUV is attenuated by
dust; even though the heating rate remains high, the cooling rate is
enhanced by the presence of molecules.  As a result, the temperature
in the atomic heated surface layer drops to more moderate values at a
column density $N_H < N_{FUV}$, and then at $N_H = N_{FUV}$ the
temperature drops further because of the loss of heating.

The dust abundances, and hence the dust opacities, affect all of the
above considerations. St\"orzer \& Hollenbach (1999) modeled 10
proplyds in Orion and found that the best fit FUV dust cross section
per H nucleus in these photoevaporative flows was approximately
$\sigstar$ = 8$\times 10^{-22}$ cm$^2$ per H nucleus, which is about
0.3 times the value for standard interstellar dust. This finding
provides strong evidence for moderate coagulation and settling of the
dust in the surfaces of these disks at distances $r\sim 30-100$ AU
from the central star (for disk ages $\sim 1$ Myr).  We use this value
of the FUV dust opacity for the models presented in this paper.  We
further discuss the coagulation of the dust in disks later in \S5,
where we apply our models to the formation of Kuiper Belt Objects and
the possible presence of a sharp cutoff in Kuiper Belt Objects beyond
50 AU in the solar system.

\subsection{Gas Cooling Mechanisms} 

The PDR code includes a large number of cooling mechanisms (see
Tielens \& Hollenbach 1985, Kaufman et al. 1999).  For most of the
cases considered in this paper, where gas densities $n \approx 10^4 -
10^8$ cm$^{-3}$ and $T \approx 100-3000$ K, the most important gas
coolants include gas-grain collisions (the grains are typically much
cooler than the gas, of order $T = 10 - 50$ K), and collisional
excitation followed by radiative decay of [CII] 158 $\mu$m, [OI] 63
$\mu$m, and the rotational and vibrational transitions of H$_2$, CO,
and OH.  In rough terms, cooling by adiabatic expansion is only
important when the flow time scale is relatively short compared to the
heating time. In this setting, the flow is slow (subsonic) near the
disk and supersonic in the outer region. We treat this outer regime by
assuming that the flow has a constant temperature and constant flow
speed for $r > r_s$.

\subsection{Thermal Balance}

The PDR code assumes thermal balance, i.e., that the sum of the
heating rates is equal to the sum of the cooling rates. Given enough
time, gas will reach this state of thermal balance. In this context,
however, the gas is flowing outward and has a limited time to reach
its preferred thermal state.  To justify the assumption of thermal
balance, we must compare the flow time with the heating time. The time
scale $t_f$ for the flow to cross a radial scale $r$ is given by 
\be
t_f \equiv r/v_f \simeq 95 \left( {r \over {20\ {\rm AU}}} \right) 
\left( {v_f \over {1\ {\rm km/s} } } \right)^{-1} \ {\rm years},
\ee
where $v_f$ is the flow speed. The heating timescale $t_h$ is given by 
\be
t_h \equiv nkT / \heat \simeq 
4.4 \left({n\over {10^6\ {\rm cm}^{-3}}}\right)
\left({T \over 10^3\ {\rm K}}\right)
\left( {\heat \over {10^{-15} \ {\rm erg \ cm^{-3} \ s^{-1}} } } 
\right)^{-1} \ {\rm \ years} , 
\ee
where $\heat$ is the heating rate per unit volume. To obtain the
benchmark value of the heating rate, we have used a number density $n
= 10^6$ cm$^{-3}$ and a radiation field with $G_0$ = 3000; keep in
mind that $\cal H$ depends on both $n$ and $G_0$.  For this case, 
the heating rate $\heat \simeq 10^{-15}$ erg cm$^{-2}$ s$^{-1}$ and
for hot ($T \simeq 1000$ K) surface gas, and we find that $t_h < t_f$.
As a result, thermal balance is justified in this fiducial case. We
expect that the gas will be close to thermal balance in the inner
regions where the flow is subsonic. In the outer region, however, the
flow becomes supersonic and the heating time can exceed the dynamical
flow time scale. In this paper, we take into account this effect by
assuming that the gas temperature becomes isothermal in the outer
region $r \ge r_s$. Specifically, we allow the temperature to increase,
according to the heating/cooling treatment described here, from the
inner boundary $r_d$ out to the sonic point $r_s$, and then assume
that the temperature remains constant for $r > r_s$.

\subsection{Chemistry}

The PDR code includes 46 chemical species and 222 chemical reactions
(Tielens \& Hollenbach 1985, Kaufman et al. 1999). The code is
designed to follow the dominant coolants and includes substantial
hydrogen, oxygen, and carbon chemistry, but only limited sulfur
chemistry and essentially no nitrogen chemistry. The code calculates
the steady state molecular abundances as a function of column density
$N_H$ (or, equivalently, the visual extinction $A_V$). The molecular
chemistry is quite complicated, but has been elucidated in previous
work (again, see Kaufman et al. 1999). In the present context, the
most important chemical activity is that the H$_2$ abundance rises
substantially once the column density from the disk surface exceeds
$N_H \simeq 10^{20}$ cm$^{-2}$. This formation of molecular hydrogen
is the precursor to the formation of other molecules and it thus marks
the onset of molecular cooling, which in turn leads to a drop in the
gas temperature $T$ at large column density.

In the outflowing gas, the predominant chemistry involves the
formation of molecules and their destruction by photodissociation.
The chemical time scales for most molecules is comparable to the
photodissociation time 
\be
t_{\rm chm} \approx 0.1 \Bigl( {G_0 \over 3000} \Bigr)^{-1} \, 
{\rm year} \, , 
\ee
which is short compared to the flow time scale, thereby justifying the
assumption of steady state. Although this time scale applies to most
molecules, H$_2$ is an important exception. Because H$_2$ self
shields, its photodissociation time scale is much longer. As a result,
H$_2$ can advect closer to the surface than the steady state model
predicts (e.g., Bertoldi \& Draine 1996, St{\"o}rzer \& Hollenbach
1998).  The extra H$_2$ increases the heating rate as well as the
cooling rate.  Although this effect could be significant, a detailed
treatment is beyond the scope of this paper because it requires a
time-dependent PDR code coupled with a hydrodynamical calculation. The
most important ramification of this effect is its signature in the
temperature structure of the flow ($T$ as a function of column density 
$N_H$). However, the results of St{\"o}rzer \& Hollenbach (1998)
indicate that the effects are relatively modest, so this paper still
provides a good first approximation to the photoevaporation problem
for $r_d < r_g$.

\subsection{Temperature Profiles} 

The net result of the PDR code calculations is a determination of the
gas temperature as a function of both number density $n$ and visual
extinction $A_V$ (or column density $N_H$ = $\delta_{UV} A_V/\sigstar$, 
where $\delta_{UV} = 1.8$ is the conversion factor between visual
extinction and FUV optical depth). The resulting temperature profiles
are shown in Figure 2. Each panel corresponds to a given external
radiation field. The temperature profiles, plotted as a function of
$A_V$, are shown for each number density. Recall that the PDR code
itself uses a constant density, plane parallel configuration and
calculates the temperature by using a detailed treatment of heating,
cooling, and chemistry.  The key assumption in this work is that the
resulting temperature dependences $T(n,A_V)$ provide a good working
approximation for other geometrical configurations, in particular the
radial flow fields that arise during the photoevaporation process (see
\S 3 and 4).

Examination of the temperature profiles in Figure 2 shows a number of
significant trends. For a given radiation field and a given density,
the temperature approaches a nearly constant value at low visual
extinction $A_V$. The temperature decreases slowly with increasing
$A_V$ until the optical thickness of the dust becomes significant at
$A_V \sim 1$. The temperature then decreases sharply as $A_V$
increases further. The outer temperature (at low $A_V$) generally
increases with the intensity $G_0$ of the radiation field (as
expected), although this outer temperature also varies significantly
with the density. Finally, the temperature is not a smooth monotonic
function of visual extinction $A_V$ or number density $n$. The
temperature profiles exhibit a great deal of structure, mostly due to
the heating and cooling effects of various molecules (and atoms) that
come in and out of existence with varying $n$ and $A_V$.
 
\section{PHOTOEVAPORATION OF SUPERCRITICAL DISKS}  

In this section, we review the simple models for photoevaporation that
have been developed previously for the supercritical regime where $r_d
> r_g$ (e.g., Johnstone et al. 1998). In this regime, the disk gas
that resides at radii from $r_d$ to $r_g$ has little gravitational
binding energy and readily achieves supersonic flow speeds near $r
\sim r_d$.  Because the disk edges have less surface area than the
faces, most of the mass loss is driven off of the top and bottom disk
surfaces (in contrast to the case of subcritical disks -- see \S 4).
These disks generally have surface density profiles that decrease with
radius, and the disks shrink from the outside inwards as they
evaporate in external FUV fields (this is also the case for
subcritical disks).

The vertical flow off the disk follows pressure gradients that rapidly
turn the flow into the radial direction by the time the flow reaches
$r \gta r_d$. Previous studies of this supercritical flow have
generally worked in the limit $G_0/n \ge 10^{-2}$ cm$^3$ where the
flow column density $N_H \approx N_{FUV} = N_C$ (again, see Johnstone
et al.  1998). In this regime, the condition $\tau_{FUV} \sim 1$
defines the total column density so that the outgoing flow itself is
the limiting factor: As the mass outflow rate $\dot M$ increases, the
optical depth of the flow increases and the $\tau_{FUV} = 1$ limit is
reached.  If the flow were to increase beyond this level, the flow
would become so optically thick that the driving FUV photons could no
longer penetrate down to the disk.

These models also assume that the flow speed approaches a constant
value in the region where most of the column density resides. With
this assumption, in conjunction with radial symmetry, the continuity
equation implies that the density field of the flow takes the form 
\be
n (r) = n_b (r_d/r)^2 \, ,
\ee
where $n_b$ is the number density at the base of the flow.
The total column density $N_H$ is given by the integral
\be
N_H = \int_{r_d}^\infty n (r) \, dr \, = n_b r_d \, .
\ee
The dust optical depth is given by $\tau_{FUV} = \sigstar N_H$, where
$\sigstar \approx \ 8\times 10^{-22}$ cm$^2$ is the appropriate cross 
section for dust grains interacting with FUV radiation (e.g., see SH99). 
The optical depth unity surface thus defines a constraint on the base 
density $n_b$, i.e., 
\be
n_b r_d = \sigstar^{-1} \approx 10^{21} {\rm cm}^{-2} \, .
\ee

For the case of a large disk with $r_d > r_g$, the
mass outflow rate is given by
\be
{\dot M} = {\cal F} 4 \pi r_d^2 n_b a_S \muc \, ,
\ee 
where $\cal F$ is the fraction of the solid angle subtended by the
outflow and $\muc$ is the mass of the gas molecules (the conversion
factor between number density and mass density). For $r_d > r_g$, the
flow from the disk surface and the disk edge merge at roughly $r_d$ to
$2r_d$, creating a nearly spherically symmetric flow so that ${\cal F}
\sim 1$. Using the $\tau_{FUV} = 1$ constraint to define the value of
the base density $n_b$, we obtain an estimate of the mass loss rate, 
\be
{\dot M} = 4 \pi {\cal F} \muc \sigstar^{-1} a_S r_d \, 
\approx 1.2 \times 10^{-7} M_\odot {\rm yr}^{-1} {\cal F} 
\Bigl( {a_S \over 2 \, {\rm km/s}} \Bigr) 
\Bigl( {r_d \over 100 \, {\rm AU}} \Bigr) \, , 
\label{eq:mdotzero} 
\ee
where everything is specified except for the disk radius $r_d$ and the
sound speed $a_S$ of the flow ($a_S$ is set by the temperature, which
is set by the external radiation flux).  In the second approximate
equality, we have defined a benchmark evaporation rate for the
supercritical regime using $r_d$ = 100 AU and $a_S$ = 2 km/s (for $T
\approx$ 600 K). When a typical solar nebula (with disk mass $M_d$
= 0.03 $M_\odot$) experiences mass loss in this supercritical regime,
the evaporation time scale is only about 0.25 Myr, much less than 
the expected time scale for giant planet formation ($t \sim10$ Myr;
e.g., Lissauer 1993).  Thus, supercritical evaporation can readily
evaporate nebular disks and compromise the planet formation process in
the outer regions.  However, for $M_\ast$ = 1.0 $M_\odot$ and $T$ =
600 K, the critical radius $r_g$ = 160 AU so that only the largest
disks can experience supercritical mass loss.  Many disks will live in
the subcritical regime, and we must generalize this treatment, as
outlined in the following section.

\section{PHOTOEVAPORATION OF SUBCRITICAL DISKS}  

In this section, we generalize the photoevaporation model to include
cases where the disk radius is smaller than the critical radius, i.e.,
$r_d < r_g$. In this regime, the disk material is not immediately free
to escape because the sound speed in the outer layer (that heated by
FUV radiation) is still less than the escape speed. However, the disk
has an atmosphere that extends beyond the nominal radius $r_d$ and
some portion of that atmosphere will extend above the $r=r_g$ surface
and can be susceptible to evaporation. As material leaves the system,
an outward flow develops. The result is much like a Parker Wind
solution: The flow starts subsonically at $r \ll r_g$, accelerates up
to a sonic point at $r_s \lta r_g$, and then expands supersonically
outwards.  We thus need to make a simple model of the disk atmosphere
and the accompanying flow.

\subsection{Basic Flow Geometry} 

This problem contains four important length scales. In the
supercritical regime considered previously (\S 3), the escape radius
$r_g$ (eq. [\ref{eq:rcrit}]) marks the inner boundary of the flow. In
the subcritical regime considered here, the inner boundary of the flow
is the disk radius $r_d$ where the flow speed is subsonic ($v \ll a_d$). 
As the material flows outwards, the Mach number increases and the flow
eventually exceeds the local sound speed at a sonic point $r_s$ (which
marks the outer boundary of our numerical calculations).  The sonic
point $r_s$ is smaller than (but roughly comparable to) the escape
radius $r_g$. The final length scale of interest is the disk scale
height $H_d$ at $r_d$. If the disk extended out to the escape radius, 
then $H_d \sim r_d \sim r_g$; in the subcritical regime, however, the
disk is relatively ``thin'' so that $H_d < r_d$. As a result, as shown
in Figure 1, the basic length scales for these evaporating disks obey
the ordering 
\be 
H_d < r_d < r_s < r_g \, . 
\label{eq:ordering} 
\ee 

The disk atmosphere behaves differently in the radial ($\hat r$) and
vertical ($\hat z$) directions. As shown below and in the Appendix,
the outflow from the disk edges (the radial flow) dominates the
outflow from the disk faces (the vertical flow). As a result, we can
assume that the essential part of the outflow takes place radially
outwards from the disk edges. With this simplification, we construct a
quasi-spherical disk model and take into account the fraction $\cal F$
of the solid angle that is subtended by the outflow from the disk edge
(for a given scale height $H_d$). We also assume that radiation can
hit the system at any angle, so that the disk receives its full quota
of FUV radiation. A schematic diagram of this system is shown in
Figure 1.  Although the vertical flow is secondary in importance (for
determining the outflow rate), the polar regions are not evacuated;
these cavities will be filled by (more slowly moving) material that
will attenuate the incoming FUV radiation.

To show that the radial flow tends to dominate the vertical flow, we
consider the (one dimensional) profiles of density in the two
directions in the hydrostatic limit. The density profile in the
vertical direction can be written in the form 
\be 
\log n/n_d = - {G M_\ast \over r_d a_d^2} \Bigl[ 1 - 
{1 \over 1 + z^2 / r_d^2} \Bigr] \, , 
\label{eq:vertpro} 
\ee
where we use an isothermal approximation so that $a_d$ is the
isothermal sound speed at the disk surface. Note that this form does
not assume small $z \ll r$. However, this equation is strictly valid
for only the outermost annulus of the disk (at smaller disk radii, the
gas is deeper in the potential well and contributes little to the mass
outflow).

Similarly, we can integrate the hydrostatic force equation in the
radial direction to obtain 
\be 
\log n/n_d = - {G M_\ast \over 2 r_d a_d^2} 
\bigl( 1 - {r_d \over r} \bigr)^2 \, . 
\ee  
Since the vertical coordinate $z$ starts at $z=0$, and the radial 
coordinate starts at $r=r_d$, we define $x=r - r_d$ and rewrite 
the radial profile in the form 
\be 
\log n/n_d = - {G M_\ast \over 2 r_d a_d^2} 
\bigl( {x \over r_d} \bigr)^2 \, 
\bigl( {1 \over 1 + x/r_d} \bigr)^2 \, . 
\label{eq:radpro} 
\ee  
A straightforward comparison shows that the right hand side of
equation (\ref{eq:vertpro}) is always greater than [or equal to -- but
only at the disk surface ($z=0$) at $r_d$ (where $x=0$)] the right hand
side of equation (\ref{eq:radpro}).  This result implies that the
effective scale height of the density profile is always larger in the
radial direction than in the vertical direction (so that the density
falls off more slowly in $r$). Because the density tends to decrease
more quickly in the vertical direction than in the radial direction,
the density remaining at the sonic point will be greater for radial
flow, and the mass loss rate will be larger for radial flow. In this
paper, we thus assume that the radial portion of the flow dominates,
and model the system using a quasi-spherical calculation. 
We also note that equation (\ref{eq:radpro}) for the radial density
profile holds over the entire area of the disk edge, whereas equation
(\ref{eq:vertpro}) for vertical flow is only valid only for the
outermost annulus of the disk (at $r_d$). At smaller radii, the gas is
deeper in the potential and less likely to get out. As a result, the
area of the disk edge has a greater working surface area than the disk
face. This area argument thus argues that the radial flow from the
disk edge is most important (see the Appendix for a more quantitative
argument showing that radial flow dominates the vertical flow for $r_d
\ll r_g$).

For these systems where the radial flow from the disk edge dominates
the vertical flow from disk faces, we calculate the mass outflow rates
by constructing a radial wind solution, but assume that only a
fraction $\cal F$ of the solid angle is filled by the flow. The disk
edges are essentially a cylinder of radius $r_d$ and height $2H_d$,
and thus subtend a given fraction of the $4\pi$ steradians of solid
angle centered on the star. Since the outflow is nearly radial, the
solid angle subtended by the flow remains constant with radius and is
given by 
\be
{\cal F} = {H_d \over (H_d^2 + r_d^2)^{1/2} } \, , 
\ee
where the disk scale height $H_d \approx r_d a_d (GM_\ast/r_d)^{-1/2}$.
We note that the remaining solid angle is not evacuated. These regions
contain slower moving material that will contribute to the attenuation 
of incoming FUV radiation, but will contribute relatively little to
the total mass loss rate.

\subsection{The Outer Region} 

The flow in the outer region, beyond the sonic point, provides an
outer boundary condition for the flow in the region of interest ($r_d
\le r \le r_s$). In the outer region where $r > r_s$, we assume the
the flow is radial and has constant flow velocity. In other words,
beyond the sonic point we assume that the flow has the same properties
as found for the supercritical regime (\S 3), albeit with lower mass
loss rates $\dot M$. As a result, the density takes the form 
\be
n_{out} = n_s (r_s/r)^2 \, , 
\ee
where $n_s$ is the number density at $r_s$. The column density \
$N_{s \infty}$ of the outer region is given by the integral 
\be
N_{s \infty} = \int_{r_s}^\infty n_{out} (r) \, dr \, = n_s r_s \, , 
\ee  
and the corresponding dust optical depth of this region is given by 
\be
\atauess = \sigstar n_s r_s \, .  
\ee 

\subsection{Basic Equations of Motion} 

If we include rotation for the force balance in the circumstellar 
disk, the radial force equation takes the form 
\be 
v {dv \over dr} + {1 \over \rho} {dP \over dr} + 
{GM_\ast \over r^2} - {j^2 \over r^3} \, = 0 \, , 
\label{eq:forcezero} 
\ee
where $j$ is the specific angular momentum. We can specify the angular 
momentum by requiring it to be the Keplerian value at the outer disk 
edge so that $j^2 = G M_\ast r_d$. Now we assume an ideal gas law for 
the pressure, i.e., $P = n k T$. To simplify the equations, 
let $\xi = r/r_d$, $f = T/T_d$, $g = n/n_d$, and $u=v/a_d$ 
(where $a_d$ is the sound speed at the disk edge). 
The force equation becomes 
\be 
u {du \over d\xi} + {1 \over g} {d\over d\xi} (gf) + 
\beta {\xi - 1 \over \xi^3} = 0 \, , 
\label{eq:force} 
\ee
where we have defined  
\be
\beta \equiv {G M_\ast \muc \over k T_d r_d} = {G M_\ast \over 
r_d a_d^2 } \, . 
\ee 
The parameter $\beta$ can also be written in the form $\beta = r_g
T_g/r_d T_d$ and thus provides a measure of how subcritical 
the disk edge is. 

Now we introduce the continuity equation, which takes the form  
\be 
{\dot M} = 4 \pi r^2 {\cal F} \muc n v \, = \, {\it constant} \, , 
\ee
where we have included the filling factor $\cal F$. 
If we define a constant $C$ according to 
\be 
C \equiv \Bigl( { {\dot M} \over 
4 \pi r_d^2 {\cal F} \muc n_d a_d} \Bigr)^2 
= {v_d^2 \over a_d^2 } \, , 
\ee 
the dimensionless form of the continuity equation becomes 
\be 
\xi^2 g u = \sqrt{C} \, . 
\ee  
We can use the continuity equation to eliminate the flow speed $u$
from the differential equation (\ref{eq:force}), and thereby obtain a 
single differential equation (in density $n$ or $g=n/n_d$) to describe
the flow.  Alternately, we can use the continuity equation to
eliminate the density from the force equation and obtain a
differential equation for the flow speed. The former approach allows
us to solve for the density structure of the flow.  The latter
approach defines the sonic point, which is necessary to define
boundary conditions. So we follow both approaches. 

\subsection{The Resulting Flow Equations} 

By eliminating the flow speed $v$ through the continuity equation, 
the force equation becomes  
\be 
{d\over d\xi} (gf) + \beta g {\xi - 1 \over \xi^3} + 
{C \over \xi^2} {d \over d\xi} \bigl( {1 \over g \xi^2} \bigr) 
= 0 \, , 
\ee
where the constant $C$ is defined above. 
It is useful to expand this equation to obtain the form 
\be 
{dg \over d\xi} \Bigl(f - {C \over \xi^4 g^2} \Bigr) = 
{2C \over \xi^5 g} - \beta g {\xi - 1 \over \xi^3} - 
g {df\over d\xi} \, . 
\label{eq:dense} 
\ee 

The PDR models specify the temperature as a function of column density
$N_H$ or, equivalently, the visual extinction $A_V$.  Here we can work
in terms of the variable $\atau \equiv N_H \sigstar$, which is a dust
optical depth.  To complete the specification of the problem
(essentially, in order to determine the temperature), we need to
include the differential equation that determines the optical depth
$\atau$ as a function of $\xi$. In dimensionless form, $\atau$ is
determined by the equation 
\be  
{d \atau \over d \xi} = - \sigstar r_d n_d g = - \tau_d g \, ,
\label{eq:adef} 
\ee
where the second equality defines $\tau_d = \sigstar r_d n_d$. 

Notice that the evaluation of $df/d\xi$ is a bit subtle, since $f$
(the dimensionless temperature) is a function of both the density
($g$) and the column density (or $\atau$). Thus, we can write 
\be 
{df \over d\xi} = {1 \over T_d} {\partial T \over \partial n} 
{dn \over d\xi} + {1 \over T_d} {\partial T \over \partial \atau} 
{d\atau \over d\xi} \, . 
\label{eq:fprime} 
\ee 
Notice, however, that the definition of $df/d\xi$ contains the
derivative of the density ($dn/d\xi$ or $dg/d\xi$) so that equation
(\ref{eq:dense}) remains in implicit form. 
 
In this approach, the stellar mass $M_\ast$, radius $r_d$, and outer
disk temperature $T_d$ are given system parameters, so the constant
$\beta$ is specified. However, the density at the disk edge $n_d$ and
the constant $C$ which determines the flow velocity are not determined
in advance. The density $n_d$ enters into the problem by pinning down
the scale for the column density. The constant $\sqrt{C}$ is the
dimensionless mass outflow rate (i.e., $\dot M$), the quantity that we
want to calculate in the end. For given estimates of $n_d$ and $C$,
the differential equations (\ref{eq:dense}) and (\ref{eq:adef}), along 
with the definition (\ref{eq:fprime}), can be integrated outwards to 
the sonic point. 

\subsection{Specification of the Sonic Point} 

To determine the location of the sonic point, we need to eliminate the
density from the force equation (instead of the velocity) by using the
continuity equation. Using the same dimensionless formulation as before, 
we obtain 
\be 
{1 \over u} {du \over d\xi} \bigl(u^2 - f \bigr) = 
{2 f \over \xi} - {df \over d\xi} - \beta {\xi - 1 \over \xi^3} \, . 
\label{eq:uforce} 
\ee 
At the sonic point, $u^2 = f$, the left hand side of the equation
vanishes, and so the right hand side of the equation must vanish
also. This constraint implies the relation 
\be 
2 f \xi^2 - \beta (\xi - 1) - \xi^3 {df \over d\xi} = 0 \, , 
\label{eq:cubic} 
\ee
which thereby defines the sonic point. The full definition of the
sonic point thus involves a cubic equation in $\xi$. In practice,
however, the final term is relatively small. Furthermore, because the
flow time must be longer than the heating time in order for the
outflowing gas to change its temperature, the gas tends to become
isothermal near the sonic point so that $f \to$ {\sl constant}. As
outlined above (see \S 2), we assume that the flow reaches both a
constant temperature and a constant flow speed in the outer region.
With this specification of our outer boundary condition, the sonic 
condition is the solution to equation (\ref{eq:cubic}), which becomes 
quadratic in the limit $df/d\xi$=0, i.e.,  
\be
\xi_s = {\beta \over 4 f} \Bigl[ 1 + 
\bigl( 1 - 8 f/\beta \bigr)^{1/2} \Bigr] \, . 
\label{eq:sonic} 
\ee
We have chosen here the physically realistic root of the quadratic
equation, i.e., the root that has the form $\xi_s \to \beta/2f$ in the
limit of large $\beta$.  The other root approaches unity in this limit
and is unphysical.

\subsection{Iteration Procedure} 

If we specify the radiation field $G_0$, the disk size $r_d$, the
outer disk temperature $T_d$, and the stellar mass $M_\ast$, then we
need to solve self-consistently for the density $n_d$ at the base of
the flow (our inner boundary) and the constant $C$ that sets the flow
speed at the inner boundary or, equivalently, the dimensionless mass
loss rate. In the absence of external radiation, the disk would have a
density set by its temperature and surface density; the quantity $n_d$
is the density at the base of the outflow, which does occur at $r \sim
r_d$, but may be a scale height or so above the original (not
externally heated) disk itself.  For a given $G_0$, $n_d$, and $T_d$,
the PDR models gives us the value of column density or optical depth
(at FUV wavelengths) at the disk edge. This specification acts as the
inner boundary condition for equation (\ref{eq:adef}).

We don't know ({\it a priori}) the density $n_d$ at the inner
boundary, so we guess the value and invoke a constraint.  By applying
this constraint through an iterative procedure, we can converge on the
correct value. The constraint that we invoke is that our solution must
match onto the flow in the outer region beyond $r_s$. In this outer
region, as outlined above, we assume a steady flow with $n \sim
r^{-2}$ and hence the column density from $r_s$ to $\infty$ is given
by $N_{s \infty} = n_s r_s$, or equivalently, $\atauess = \sigstar n_s
r_s$. Our solution to the differential equation defines when we get to
the sonic point $\xi_s$, where the dust optical depth will 
have a calculated value $\atau(\xi_s)$. Thus, at the sonic point $\xi_s$,
we know the value $\atau(\xi_s)$. But we also know the derivative
$d\atau/d\xi$ and hence we know $n_s$ and also $\atauess$. In general, 
the value $\atau(\xi_s)$ will not be equal to the correct value 
$\atauess = \sigstar n_s r_s$ = $\tau_d \xi_s g(\xi_s)$. In other 
words, we are searching for a zero of the function 
\be
F_\tau (n_d) = \atau(\xi_s) - \tau_d \xi_s g(\xi_s) \, , 
\ee
where $\xi_s$ is the sonic point defined previously and where
$\atau(\xi_s)$ and $g(\xi_s)$ are the calculated solutions to the
differential equations (\ref{eq:dense}) and (\ref{eq:adef}). If the
function $F_\tau$ is not zero, then we can go back and choose a new
estimate for $n_d$ (or $d\atau/d\xi$ at $\xi=1$) and integrate
outwards again, and then repeat until we converge upon $F_\tau = 0$
[$\atau(\xi_s) = \atauess$].  In principle, we can carry out this 
iteration procedure for any value of the other unspecified constant
$C$. Notice that the hydrostatic approximation is equivalent to
assuming that $C=0$.

%In practice, however, if the estimated value of $C$ is too large, the
%procedure won't converge. The reason for this difficulty, briefly, is
%that the value of $C$ determines, in part, the location of the sonic
%point, and the differential equation is not defined as the flow passes
%through the sonic point. Mathematically speaking, the flow fields are
%continuous, but their derivatives are discontinuous at the sonic point
%(i.e., the functions are $C^0$ but not $C^1$ at $\xi = \xi_s$).

The value of $C$ specifies the mass outflow rate in that $\sqrt{C}$ 
is the dimensionless mass loss rate. In other words, by definition,  
\be 
{\dot M} = 4 \pi r_d^2 {\cal F} \muc n_d a_d \sqrt{C} \, . 
\label{eq:mdotc} 
\ee 
But the mass outflow rate at the outer boundary (the sonic point) 
is given by 
\be 
{\dot M} = 4 \pi r_d^2 {\cal F} \muc n_d a_d \xi_s^2 g \sqrt{f} \, . 
\ee
Thus, for consistency, we must invoke the constraint $C$ = 
$\xi^4 g^2 f$ at the sonic point. In other words, we are searching 
for the zero of the second function 
\be
F_{ {\dot M} } = C - \Bigl[ \xi^4 g^2 f \Bigr]_{\xi_s} . 
\ee 

One way to carry out the iteration procedure is as follows: First we
estimate $C$. For that value of $C$, we estimate the density $n_d$. We
then carry out the iteration procedure on $n_d$ until it has converged
to the proper value (so that $F_\tau = 0$) for the working value of
$C$.  In general, $C$ will not have the right value to conserve mass
(to satisfy the second constraint $F_{{ \dot M}}$ = 0), so we pick a
new value of $C$. For the new value of $C$, we run the iteration
procedure on $n_d$ until it converges, and so on. When convergence is
reached, the value of the dimensionless constant $C$ determines the
total effective mass outflow rate through equation (\ref{eq:mdotc}). 

\subsection{Analytic Scalings} 

The Appendix provides an analytic solution for the photoevaporative
mass loss rate for the simple approximation that the external field
$G_0$ heats the disk surface to a constant temperature $T_s$ to a
critical depth $N_C$. Under this set of approximations, the solution
for disks with $r_d < r_g$ takes the form
\be
{\dot M} = C_0 N_C \muc a_s r_g \Bigl( {r_g \over r_d} \Bigr) 
\exp\bigl[ -r_g/2r_d \bigr] \, ,
\label{eq:analytic} 
\ee 
where $C_0$ is a dimensionless constant of order unity and where $a_s$
is the sound speed appropriate for the temperature $T_s$.  We can use
our numerical results to provide a specification of the constant $C_0$
by matching the mass loss rates for a given value of $r_d/r_g$.  This
matching procedure is sensitive to the assumed matching point because
the (numerically determined) temperature distribution does not
suddenly drop at $N_H = N_C$, but rather continuously falls with
increasing column density.

As shown in the Appendix, this scaling law for $\dot M$ does not drop
appreciably with decreasing disk size $r_d$ until the disk is
significantly smaller than the critical radius, i.e., until $r_d/r_g
\lta 0.15$. The analytic treatment also shows that the mass loss rate
from the disk surface (the vertical flow) is smaller than the mass
loss rate from the disk edges (the radial flow) by a factor of $\sim
(r_d/r_g)^{1/2}$. The flow from the disk edge thus dominates for $r_d
\ll r_g$.

\subsection{Results} 

The formulation developed thus far allows us to calculate the mass
outflow rates from circumstellar disks, as a function of stellar mass
$M_\ast$, outer disk radius $r_d$, temperature boundary conditions
$T_d$, and the intensity $G_0$ of the external radiation field.  The
resulting fluid fields for a converged model are shown in Figure 3. In
this system, a 30 AU disk surrounds a 1.0 $M_\odot$ star, and the
star/disk system is exposed to an FUV radiation field of intensity
$G_0$ = 3000. As shown in Figure 3, the flow begins subsonically at
the disk edge (with Mach number ${\cal M} \sim0.09$) and smoothly
approaches the sound speed at a radius a few times larger than that of
the disk.  The flow speed and the temperature increase outwards, while
the density decreases. All of the functions vary (nearly) as
power-laws, as indicated by the nearly straight lines on the log-log
plot. The density profile is actually a combination of power-law and
exponential behavior (see eq. [\ref{eq:analytic}]). The power-law
behavior is dominant for the regime of parameter space where
substantial flow develops; when the exponential behavior dominates,
the mass outflow rate becomes exponentially suppressed.

Figure 4 shows the mass loss rates as a function of disk radius for a
typical disk surrounding a $M_\ast$ = 1.0 $M_\odot$ star embedded in
external FUV radiation fields with $G_0$ = 300 -- 30,000. For the
central value $G_0$ = 3000, the figure shows the result for three
different choices of the temperature at the inner disk edge,
specifically $T$(30 AU) = 60 K, 75 K, and 90 K. The resulting
evaporation rates are relatively insensitive to this inner boundary
condition and we will adopt the central value, $T$(30 AU) = 75 K, as
our working `standard' value. Figure 4 shows that for a given FUV
field (which roughly fixes $N_C$ and $a_s$) the mass loss rate
decreases with shrinking disk radius $r_d$, as expected by the
analytic scaling law (eq. [\ref{eq:analytic}]) where $\dot M$ $\propto
\exp[-r_g/2r_d]$. Our numerical approach contains approximations that
do not allow us to find solutions for large $r_d/r_g \gta 0.14$
(because the sonic point solution of eq. [\ref{eq:sonic}] becomes
complex). Furthermore, the mass loss rate $\dot M$ drops rapidly with
smaller $r_d/r_g \lta 0.14$ so that the gas is too dense (at the base) 
to be modeled accurately with the PDR code. Therefore, the numerical
results are confined to a relatively small range of $r_d/r_g$, but a
large range of $\dot M$. We can use our analytic approximation
(eq. [\ref{eq:analytic}]) to provide estimates for the mass loss rates
when $r_d/r_g \gta 0.14$. The result is shown in Figure 5 for the case
of $M_\star$ = 1.0 $M_\odot$ and $G_0$ = 3000, where we have specified
the dimensionless constant $C_0$ to match the numerical solution.

Figure 4 also shows that the mass loss rates $\dot M$ are a sensitive
function of the intensity $G_0$ of the FUV radiation field. For an
external radiation field with $G_0$ = 300, the evaporation rate is
almost inconsequential (for disks with $r_d \lta 100$ AU). For
stronger radiation fields with $G_0$ = 3000 -- 30,000, however, the
mass loss rates are significant. The mass loss rates are sensitive 
to $G_0$ because higher values of the radiation intensity lead to 
higher temperatures and lower critical radii $r_g$ ($\propto T^{-1}$). 
Since ${\dot M} \propto \exp[-r_g/2r_d]$ (approximately), a modest 
increase in temperature can lead to a significant increase in the 
mass loss rate. 

Figure 6 shows how the evaporation rate depends on the mass of the
parental star. All of these models use an external FUV radiation field
with $G_0$ = 3000 and assume our standard boundary conditions. The
curves show the resulting mass loss rates for stellar masses $m$ =
$M_\ast/(1 M_\odot)$ = 0.25 -- 1.0. Notice that the mass loss rate is
a sensitive function of the central stellar mass ($m$); at $r_d
\approx$ 20 AU, the evaporation rate varies by an order of magnitude
over the range of stellar masses used here. The mass loss rates are
sensitive to $M_\ast$ because the critical radius $r_g \propto
M_\ast$.  Since ${\dot M} \propto \exp[-r_g/2r_d]$, decreasing the 
stellar mass (with a corresponding decrease in $r_g$) leads to a
rapidly increasing mass loss rate $\dot M$. As we explore in greater
detail below (\S 7), this result implies that low mass stars can
easily lose the gas in their circumstellar disks.

Figure 7 shows the various length scales in the problem. Here, the
disk scale height $H_d$, the sonic radius $r_s$, and the critical radius
$r_g$ are plotted as a function of disk radius $r_d$. For this model,
the stellar mass $M_\ast$ = 1.0 $M_\odot$ and the FUV radiation field
has intensity $G_0$ = 3000. The critical radius $r_g$ depends on the
gas temperature according to equation (\ref{eq:rcrit}). In these
models, the temperature approaches a constant value in the outer
region where the flow is supersonic. For systems with larger disk
radii $r_d$, the flow reaches supersonic speeds more easily, at lower
temperature, and the critical radius ($r_g \propto T_s^{-1}$ as defined
here) increases with $r_d$.  The curves depicting both the sonic point
and the critical radius show (non-monotonic) structure, which is a
reflection of the structure in the relationship between temperature
and visual extinction (Fig. 2). As expected, the critical radius is
always much larger than the disk radius, by almost an order of
magnitude, for this regime of parameter space. The critical radius
$r_g \approx 4 r_s$ for systems with relatively high $\dot M$ as 
shown here (where the factor $(1-8f/\beta)^{1/2} \ll 1$ in eq. 
[\ref{eq:sonic}]). For smaller mass loss rates (smaller disk radii
$r_d$), $r_g \approx 2 r_s$ as noted in \S 4.4. Another measure of how
far the disks are from being supercritical is to compare the minimum
value of $r_g$ (corresponding to the highest temperature accessible
for a given radiation field) with the disk size $r_d$. This minimum
$r_g$ is about 160 AU for the case shown here. Nonetheless, as shown
in the previous figures, the disk experiences significant mass loss in
this subcritical state. Notice also that the length scales obey the
ordering of equation (\ref{eq:ordering}).

\subsection{Evaporation Time Scales} 

To convert our results into time scales for disk evaporation, we need
to account for the mass supply in the disk. Here we assume that the
surface density is given by the simple power-law form 
\be
\Sigma (r) = \Sigma_0 \Bigl( {r_0 \over r} \Bigr)^p \, ,
\label{eq:surfdensity} 
\ee
where $r_0$ is the initial outer radius of the disk and $\Sigma_0$
is the corresponding outer surface density. The coefficient
$\Sigma_0$ is determined by the total starting disk mass, i.e.,
\be
M_{d0} = {2 \pi \over 2-p} \Sigma_0 r_0^2 \, ,
\ee
where we have made the approximation $r_0 \gg R_\ast$ (the stellar
radius, or inner disk radius).  As the disk evaporates, we assume that
mass loss occurs from the outside to the inside; specifically, we
assume that all of the mass is evaporated from a given annulus before
the mass loss moves inward. The disk mass as a function of time is
then given by 
\be
M_d (r_d) = M_{d0} \Bigl( {r_d \over r_0} \Bigr)^{2-p} \, ,
\ee 
where $r_d$ is the time-dependent disk radius ($r_d < r_0$).
For the sake of definiteness, we take $p=3/2$ throughout this paper, 
and normalize the surface density such that 
\be 
M_d (r_d) = 0.05 M_\ast \Bigl( {r_d \over 30 \, {\rm AU} } 
\Bigr)^{1/2} \, . 
\ee 
Notice that this formula remains valid for disk radii 
$r_d > 30$ AU.  The evaporation time $\tevap$, for a given 
disk radius, is thus given by 
\be
\tevap = { M_d (r_d) \over {\dot M} (r_d) } \, .
\ee 

\subsection{Coupling of Photoevaporation and Disk Accretion} 

In addition to photoevaporation from its outer edges, the disk will
also experience disk accretion as long as it has an internal source of
viscosity. Since the disk is finite, material cannot move inwards at
all radial locations. In particular, the outer disk edge will expand
outwards on the diffusion time scale given by 
\be 
\tau_{diff} = r_d ^2 / \nu_d \, , 
\label{eq:diffuse} 
\ee 
where $\nu_d$ is the viscosity. The evaporation time scale decreases
with disk radius, whereas the disk diffusion time increases with disk
radius. In other words, as photoevaporation takes place and the disk
shrinks, the time scale required for photoevaporation grows longer,
but the time scale for the disk to replenish the mass supply (through
accretion and spreading) grows shorter. The disk will thus obtain a
quasi-equilibrium state in which the time scales for photoevaporation
and disk accretion are in balance. The disk will thus maintain a fixed
radius for as long as both processes are effective.

We can estimate the disk radius at which photoevaporation and disk
accretion are balanced. The evaporation time scales are the main focus
of this paper.  The diffusion time scale is given by equation
(\ref{eq:diffuse}), where the viscosity can be written in terms of an
`alpha prescription' via 
\be
\nu_d = {2 \over 3} \alpha a_d H_d \, , 
\ee
where $\alpha$ is the usual viscosity parameter and where we evaluate 
the sound speed and scale height at the outer disk edge. In Figure 8 
we show the resulting disk accretion time scales along with the 
photoevaporation time scales calculated in this paper. The results 
are shown for a star with mass $M_\ast$ = 1.0 $M_\odot$ with a disk 
exposed to an FUV radiation field with $G_0 = 3000$. We assume that 
the disk mass $M_d = 0.05 M_\odot (r_d / 30 {\rm AU})^{1/2}$.
Although the photoevaporation time scale is relatively insensitive to
the disk temperature at the outer edge (which represents the inner
boundary condition to the outflow problem), the disk accretion time is
more sensitive. Figure 8 shows the results for three choices of
temperature scale, $T_d (r_d = 30 {\rm AU})$, as labeled.

Disks are expected to form with a radius $r_d \sim 100$ AU, somewhat
larger than the crossover radii shown in Figure 8.  In the long term,
the disk radius will shrink down to the size at which photoevaporation
and disk spreading (from accretion) are in balance. This state should
be an equilibrium: If the disk radius were to grow, photoevaporation
would win over disk spreading and the disk would decrease its size. If
the disk became too small, then the photoevaporation would become much
less effective but disk accretion would replenish the material in the
vacated region. The disk will thus maintain this equilibrium size as
viscosity drains material onto the star and photoevaporation drains
material outward into the interstellar medium. This balance will
continue until the disk surface density becomes so small that angular
momentum transport (accretion) is no longer effective (see also 
Clarke et al. 2001, Matsuyama et al. 2003ab).  

\section{GAS REMOVAL FROM THE EARLY SOLAR NEBULA} 

The previous sections provide a working formulation to calculate
evaporation rates from circumstellar disks embedded in external FUV
radiation fields. Our first application of these results is to our own
solar system. Here, the planets Neptune and Uranus are seriously
depleted in hydrogen gas compared to solar abundances. On the other
hand, both Jupiter and Saturn are relatively gas rich. If the solar
system formed in a group or cluster environment, which in turn
provides a strong external radiation field, then gas would be lost
from the outer solar nebula through the mechanism developed above.

To fix ideas, we first assume that the disk starts off with outer
radius $r_{i}$ = $r_g$. Given that $r_g$ has a typical size of 100 AU 
(for the temperatures produced by a cluster radiation field), 
this assumption is quite reasonable. The initial mass loss rate is
then given by equation (\ref{eq:mdotzero}), the mass loss rate for a 
supercritical disk. The starting time scale $t_i$ for the disk to 
change its mass content is thus given by 
\be
t_i = {M_{di} \over {\dot M}_i} =
{a_S \sigstar M_{di} \over 4 \pi \muc G M_\ast} \, .
\ee
For a temperature of $T$ = 1000 K, the fiducial mass loss rate is
about ${\dot M}_i \approx 10^{-7}$ $M_\odot$ yr$^{-1}$.  For a
relatively large starting disk mass of $M_{di}$ = 0.1 $M_\odot$, e.g.,
the corresponding time scale is $t_i \approx$ 1 Myr.  As a result, if
our initial solar nebula extended out to the escape radius near 100 AU
(and if our solar system formed within a respectably large birth
aggregate so that $T \sim 1000$ K), then the nebula would evaporate
relatively quickly (at least at first) and become smaller.

Specifically, the nebula would shrink until its outer radius $r_d$
became significantly smaller than 100 AU. As a result, the solar
nebula would rapidly attain a smaller radius of $r_d \sim 30$ AU, a
smaller mass of $M_d \sim 0.05 M_\odot$. With these properties, the
nebula would be within the subcritical regime and it would then
continue to evaporate as described in \S 4. Notice that as the solar
nebula shrinks, the evaporation time scale is affected by two  
competing effects: As $r_d$ decreases, the mass loss rate gets 
smaller, which would tend to increase the evaporation time scale.
However, the disk mass decreases also, and this effect compensates to
some degree. In addition, the outer disk edge spreads outward on the
viscous diffusion time scale; if the disk has enough viscosity, then 
the nebula would maintain a quasi-equilibrium size as it evaporates 
(see Figure 8). 

The evaporation time scales for the solar nebula (a disk surrounding a
1.0 $M_\odot$ star) are shown in Figure 9, as a function of disk
radius, for three different intensities of the external FUV radiation
field. As a benchmark, a 30 AU disk with mass $M_d = 0.05 M_\odot$
embedded in a radiation field with $G_0$ = 3000 has an estimated
evaporation time scale of 14 Myr, which is somewhat larger than the
fiducial time scale of 10 Myr required for giant planet formation. If
the starting disk mass had the much lower value $M_d = 0.01 M_\odot$,
the evaporation time would be only about 3 Myr and the $G_0$ = 3000
radiation field could substantially affect planet formation.  For a
(much larger) FUV radiation intensity of $G_0$ = 30,000 (typical of
large clusters like the Trapezium), the evaporation time scale for a
30 AU disk (again with $M_d$ = 0.05 $M_\odot$) is only about 4 Myr,
comfortably less than the time scale expected for giant planet
formation. This level of radiation is able to evaporate the outer
portion of the early solar nebula and could provide an explanation for
the observed deficit of gas in the ice giants (Neptune and Uranus).
Even for this radiation field, however, the time scale for evaporation
is about 20 Myr at 10 AU (the location of Saturn's current orbit) and
much longer at 5 AU (the location of Jupiter). We conclude that the
region of the solar nebula where Jupiter and Saturn reside are
relatively safe from photoevaporation over the time scales relevant
for giant planet formation.

Substantial mass loss from the ice giant region of the early solar
nebula thus requires an intense FUV radiation field with $G_0$
$\approx$ 30,000. What type of solar birth environment is expected to
produce such a radiation field? For a benchmark distance of $r=0.2$
pc, for example, this radiation level can be provided by a single 30
$M_\odot$ star. Such large stars are exceedingly rare -- only about 1
out of 2400 stars are at least this massive, according to the standard
form of the stellar initial mass function ($f=dN_\ast/dM_\ast \sim
M_\ast^{-2.35}$; Salpeter 1955), so a large cluster like the Trapezium
is generally needed.  Nonetheless, stars of similar mass have been
invoked to provide enrichment of short-lived radiative species in the
early solar nebula (e.g., Cameron et al. 1995), although the preferred
distance from the massive star is much larger for optimal enrichment
and solar system survival (e.g., see Boss \& Foster 1998). 

In such an energetic environment, the EUV radiation fields will be
substantial and can affect the process of planet formation, in
addition to the FUV radiation considered here (e.g., SH99, Armitage
2000, Adams \& Laughlin 2001, Adams \& Myers 2001). However, the
effect of the EUV field in such energetic environments is somewhat
subtle. Even in the presence of strong EUV fields, the FUV fields can
dominate the mass loss rate by creating a neutral flow from the disk
surface which absorbs the EUV flux at an ionization front that lies at
several disk radii, far out into the supersonic region of the neutral
flow. St{\"o}rzer \& Hollenbach (1999) showed that for $r_d > r_g$ and
Trapezium-like conditions, the FUV dominates the mass loss rate for
$r_d \sim 100$ AU disks at distances of $d \approx 0.02$ pc to 0.4 pc
(where $G_0 = 10^4 - 10^6$). SH99 did not consider the case where $r_d
< r_g$. Once the disk shrinks to $r_d \lta 0.2 r_g$, however, the
neutral mass loss rate will decline, the EUV flux will penetrate to
the disk surface, and the EUV mass loss rate will take over.  We have
calculated the critical disk size $r_{\rm cr}$ at which EUV begins to
dominate $\dot M$ for flux from a Trapezium-like $\Theta^1$ Ori C at
distances such that $G_0$ = 30,000.  We find that FUV radiation drives
the mass loss until the disk radius shrinks to sizes less than about
$r_d \lta 10$ AU, at which point EUV takes over. The EUV flux
evaporates the disk from 10 AU down to 2 AU in about 30 Myr. The
evaporation time scale grows rapidly if the disk shrinks below $r_d
\lta 0.2 r_g \approx 2$ AU (where $r_g \sim 10$ AU for EUV heating,
which gives $T \approx 10^4$ K). Beyond this point, not even EUV
photons can lift the gas out of the potential well.  We have also
calculated the critical disk size for conditions in a moderate sized
cluster (where $N_\star \sim 300$ so that the largest star $M_\ast
\sim 9 M_\odot$) at distances such that $G_0 \approx 3000$. With a
lower mass star as the power source, the ratio of EUV to FUV photons
is lower. Here, the FUV dominates again, until the disk radius $r_d
\lta 10$ AU, where the mass loss rates are very small (${\dot M} \sim
10^{-11} - 10^{-10}$ $M_\odot$ yr$^{-1}$) and the evaporation time
scales are extremely long ($t \gg 100$ Myr).

To summarize this section, we find that an FUV radiation field of
$G_0$ = 30,000 can potentially explain the deficit of gas in the ice
giants in our solar system.  This level of radiation can effectively
evaporate the gas in the outer portion of the early solar nebula where
Neptune and Uranus now reside. The $\sim15$ Earth mass cores that form
there would have little gas to accrete and could thus develop into ice
giants (as observed). These same radiation levels will leave the
remainder of the solar nebula intact, with sufficient gas for giant
planet formation in the Jupiter/Saturn region of the nebula. In
general, however, we expect external radiation fields of this required
intensity to be somewhat rare. Unfortunately, a general assessment of
the probability for forming solar systems to experience such radiation
levels remains an open issue, which must be left for future work. We
also note that recent work suggests that the ice giants may have
formed at smaller orbital radii and then migrated outwards through
scattering encounters with Jupiter (e.g., Thommes, Duncan, \& Levison
1999). In this event, the mass loss mechanism considered here would
allow the ice giants to remain impoverished in gas even after they
migrate.

\section{FORMATION OF THE KUIPER BELT AND DEBRIS DISKS}  

The Kuiper Belt and the question of its formation is an interesting
astronomical issue for several reasons. First, since the Kuiper Belt
is an important part of our solar system, any complete theory of solar
system formation must account for its origin. Second, the existence
and observed structure of the Kuiper belt can be used to place
constraints on the process of solar system formation, including
properties of the solar birth environment (e.g., Adams \& Laughlin
2001). Third, debris disks are observed around many nearby stars
(e.g., Backman, Gillett, \& Witteborn 1992) and the dust in these
systems is provided by shattering collisions between planetesimals,
which (apparently) orbit about the central stars at distances $r$ = 10
-- 200 AU. In other words, debris disks contain rocky bodies that are
roughly analogous to the Kuiper Belt objects in our solar system. In
addition, at a given age, the amount of dust orbiting stars of similar
spectral type shows great variation. Some young stars show no evidence
for debris disks, whereas some older stars are accompanied by copious
amounts of debris dust. In this section, we show how photoevaporation
can affect the formation of the Kuiper Belt, in our solar system and
others. 

Specifically, the inventory of Kuiper Belt objects and their
associated debris dust will be suppressed if the primordial dust is
removed along with gas -- via photoevaporation -- during the first
million years in the life of the star/disk system (before dust has had
a chance to coagulate to significant sizes $b \sim 1$ cm).  To study
these effects, we need to estimate the critical size required for dust
grains to become entrained in the outflowing gas and the coagulation
time scale required for grains to attain that critical size. The
relative importance of photoevaporation then depends on the ratio of
the coagulation time scale (calculated below) to the evaporation time
scale (as calculated in \S 4).  Variations in this ratio will lead to
variations in the abundance of Kuiper belt objects and debris dust
later on.

The photoevaporation times vary widely and depend sensitively on the
mass and proximity of the nearest massive star. In one limit, a star
could be born within a large cluster like the Trapezium (e.g.,
Hillenbrand \& Hartmann 1998; St\"orzer \& Hollenbach 1999), where its
outer disk ($r \gta 50-100$ AU) evaporates much faster than dust can
coagulate. Later in their lives, such stars would show little or no
evidence for extended debris dust because the outer Kuiper Belt
objects could never form. In the other extreme, a star born in
isolation can produce numerous Kuiper Belt objects out to large
distances from the central star, and will exhibit extended debris dust
long after its formation. Most stars form in environments between these 
two extremes. 

\subsection{Critical Size for Dust Entrainment During Photoevaporation} 

In order for a dust particle to be entrained in flow and carried off
as the gas evaporates, two conditions must be met: (i) The drag force
of the gas moving past the dust particle must be greater than the
gravitational force on the dust particle from the central star. (ii)
The force must act over a sufficient time, long enough to enable the
dust particle to reach the escape speed from the system. For gas
flowing at the escape speed, the first condition can be written
(approximately) as 
\begin{equation}
b < {{3\Sigma _{gas}}\over {4 \rho _{gr}}}\left( {r \over {2H}}\right),
\end{equation}
where $b$ is the radius of the dust particle, $\Sigma_{gas}$ is the
gas surface density of the disk at radius $r$, $H$ is the scale height
of the gas at $r$, and $\rho_{gr}$ is the mass density of the grain
material. The second condition is roughly equivalent to the
requirement that the dust particle must encounter its own mass in gas
molecules (moving at the escape speed $v_{esc}$) for the dust grain to
be accelerated to $v_{esc}$. If a typical dust particle initially
resides halfway (in column density) between the midplane and the disk
surface, then a gas mass column 0.25 $\Sigma _{gas}$ will sweep by the
dust particle. This second condition can be written in the form  
\be
b < 3 \Sigma _{gas} / (16 \rho _{gr})  \, . 
\label{eq:condition2} 
\ee 
This second condition is the more stringent because $r/2H > 1$. As a
result, some relatively large particles can meet the first condition
and initially move outward from the star, but the gas flow past them
dwindles before they reach escape velocity and they fall back into
orbit (unless they also meet the second condition). 

Next we assume that the newly formed disk has a gas surface density 
of the form given by equation (\ref{eq:surfdensity}) with $p$ = 3/2. 
If the coagulated grain material has a density $\rho_{gr}\sim 1$ g
cm$^{-3}$, the second condition (eq. [\ref{eq:condition2}]) can be
evaluated to obtain the limit 
\be
b < 0.13 \ {\rm cm} \left({M_d \over {0.01 M_\odot}} \right) 
\left({r \over {100\ {\rm AU}}} \right)^{-1/2} \,  
\left({r_d \over {100\ {\rm AU}}} \right)^{-3/2} \, , 
\ee 
where $r$ is the radius at which we evaluate the limit and $r_d$ is
the location of the outer disk edge.  If we are concerned with dust
entrainment in the region 30 AU $<r<$ 100 AU, where Kuiper Belt
objects ultimately formed in our Solar System (and if $M_d \sim 0.01 
M_\odot$ and $r_d \sim 100$ AU), then the critical size for dust
coagulation is $b \approx$ 0.1 -- 1 cm.  Once dust particles coagulate
to larger sizes, they will remain bound to the system as the gas
evaporates. These remaining rocks can eventually form Kuiper Belt
objects and debris dust.

\subsection{Coagulation Timescales}

A standard scenario for dust coagulation in protostellar disks has
been developed (e.g., Weidenschilling 1997). During the collapse of a
molecular cloud core, gas and dust hit the forming circumstellar disk
in freefall, with typical speeds of $\sim$3 km s$^{-1}$ at 100 AU. The
dust passes through an accretion shock and comes to rest with the gas
in the upper atmosphere of the disk. At this stage, the dust particles
are mostly interstellar in size, with radii $b \la 0.1$ $\mu$m. Within
the disk, the dust has a tendency to settle to the midplane (due to
the vertical component of the stellar gravity force).  The gas is
supported by gas pressure, but the dust grains settle slowly through
the gas, at settling speed $v_{set}$, resisted by the drag force
exerted by collisions with gas molecules.  For the small particles,
the thermal speed $a \sim 0.1$ cm s$^{-1}$ is larger than the settling
speed $v_{set}$, and the dust particles remain in the upper atmosphere
until they begin to collide and coagulate.
 
A dust particle grows until its settling speed $v_{set}$ exceeds the
sound speed $a$, and the particle then begins its descent toward the
midplane. As it moves through gas and smaller dust particles, the dust
grain sweeps up smaller grains and grows larger. The dust particle
falls at its terminal speed, with the force of gravity balanced by the
gas drag. As the dust grain grows, its terminal speed increases, so
that the dust particle accelerates at first.  When the dust approaches
the midplane, however, the vertical component of the stellar gravity
diminishes, and the terminal speed decreases. As a rough estimate, the
dust particle attains settling speeds $v_{set} \sim 100$ cm s$^{-1}$
at 100 AU and grows to sizes of $b \sim 1$ mm (for disk mass $M_d \sim
0.01 M_\odot$) by the time it reaches the midplane. The descent to the
midplane takes about $300-1000$ Keplerian orbits around the star,
largely independent of the disk surface density. This result is due to
the cancellation of two opposing factors. At a given drift speed, a
higher density disk leads to greater gas drag, which slows down the
particles. However, a higher density disk also has more small dust
particles that are available for coagulation, so the particle grows
faster. The increased gravitational force offsets the increased gas
drag. The precise number of Keplerian orbits required depends on the
fluffiness (or fractal properties) of the coagulating dust particle as
it descends. As a working estimate, we assume that the descent to the
disk midplane corresponds to about 300 Keplerian orbits (see
Weidenschilling 1997).  In this scenario, the settling time controls
the rate of coagulation.  The time scale for small dust particles to
thermally coagulate in the upper atmosphere is almost always shorter
than the settling time of the somewhat larger particles that make the
descent. The coagulation time scale is thus given by 
\be 
t_{coag} \simeq 3\times 10^5 {\rm yr} 
\left({M_\ast / 1\ {\rm M_\odot} }\right)^{-1/2}
\left({r / {100\ {\rm AU}}}\right)^{3/2} \, . 
\label{eq:tcoag} 
\ee  
For the radii of interest for Kuiper belt formation, roughly 
30 -- 100 AU, the coagulation time scale is $t_{coag} \approx 
0.05 - 0.30$ Myr. 

\subsection{Comparison of Evaporation and Coagulation Time Scales}

The previous subsection argues that the dust coagulation time scale is
typically less than 1 Myr for radial locations corresponding to the
present-day Kuiper belt. In order for the evaporation time scale to 
compete with this (short) coagulation time scale, the photoevaporation 
time scale must lie in the supercritical regime. In this case, the 
time required for photoevaporation can be written 
\be 
t_{evap} \approx {M_d \sigstar \over 4 \pi {\cal F} \muc a_S r_d }  
\approx 0.6 \, {\rm Myr} {\cal F}^{-1} 
\Bigl( {a_S \over 1 \, {\rm km/s}} \Bigr)^{-1} 
\Bigl( {r_d \over 100 \, {\rm AU}} \Bigr)^{-1} 
\Bigl( {M_d \over 0.03 M_\odot} \Bigr) \, , 
\label{eq:tevap} 
\ee
where we have scaled the result compared to typical parameter 
values. Equating the coagulation time scale (eq. [\ref{eq:tcoag}]) 
with the evaporation time scale (eq. [\ref{eq:tevap}]), we find 
the constraint required for Kuiper belt formation to be compromised, 
i.e., 
\be 
\Bigl( {a_S \over 1 \, {\rm km/s}} \Bigr) 
\Bigl( {r_d \over 100 \, {\rm AU}} \Bigr)^{5/2} \, \ge \, 
2 {\cal F}^{-1} \Bigl( {M_d \over 0.03 M_\odot} \Bigr) 
\Bigl( {M_\ast \over 1.0 M_\odot} \Bigr)^{1/2} \, \approx 3 \, , 
\label{eq:kbolimit} 
\ee
where the approximate equality (on the right hand side of the
inequality) applies for the minimum mass solar nebula. Even at the
(rather large) radius of 100 AU, a sound speed of 3 km/s
(corresponding to a temperature of $\sim1360$ K) is necessary to
evaporate the disk faster than dust grains can coagulate.  For the
highest intensity FUV radiation field considered in this paper, $G_0$
= 30,000, the temperature at the $A_V=1$ surface is only about 600 --
700 K. In other words, the coagulation time remains shorter than the
evaporation time for nearly all of the expected radiation fields that
young solar systems might be exposed to.

Alternately, for a given radiation field and hence a given estimate
for the sound speed at the sonic point, one can derive a cutoff radius
for the existence of dust (by solving equation [\ref{eq:kbolimit}] for
the radius). For a radiation field of $G_0$ = 3000, e.g., the PDR
models indicate that the sound speed $a_S \approx 1.4 - 2.2$ km/s and
hence the cutoff radius is predicted to lie in the range $r_c \approx
110 - 140$ AU; for a radiation field with $G_0$ = 30,000, $a_S \approx
1.5 - 3.3$ km/s and the cutoff radius $r_c \approx 96 - 130$ AU. At
radial locations inside the cutoff radius, the coagulation time is
shorter than the evaporation time and dust successfully transforms
itself into centimeter-sized rocks (essentially gravel). Outside
$r_c$, most of the material (both gas and dust) is carried off to the
interstellar medium.  These considerations thus predict a reasonable
sharp cutoff for Kuiper Belt objects in circular orbits\footnote{We
emphasize circular orbits because Kuiper Belt objects formed at
smaller radii can attain highly eccentric orbits with large semi-major
axes through scattering interactions with giant planets (and smaller
bodies).}  beyond the radius $r_c$. Notice that this cutoff radius is
safely beyond the observed outer ``edge'' of our Kuiper belt at
$\sim50$ AU (see, e.g., Allen, Bernstein, \& Malhotra 2002, Trujillo
\& Brown 2001).

Particle coagulation proceeds to the critical size before the gas and
small dust evaporates at $r \la 100$ AU, even under extreme (e.g.,
Trapezium-like) conditions. Although the proplyds in Orion have shrunk
to $r_d \la 30$ AU in $\sim 0.3$ Myr (e.g., SH99), this coagulation
model predicts that a significant fraction of the dust will have
quickly coagulated into particles with $b \ga 1$ cm. These large dust
grains (rocks) will still reside in the disk, from the current disk
radius of $r_d \la 30$ AU (the boundary for gas and small dust
particles) out to about 100 AU. These rocky bodies have little optical
opacity, but are available to form Kuiper Belt objects.  Although
definitive models of planet formation are not yet available, the
accumulation of these dust grains into planetesimals (and planets)
must proceed differently in the outer disk (with no gas) and the inner
disk (within $r\sim$30 AU where gas is retained much longer).

We conclude this section with two important caveats, which could
significantly change the estimated radius $r_c$, as set by the
intersection of the coagulation time with the evaporation time. First,
the evaporation time scale is proportional to the disk mass, whereas
the coagulation time scale is almost independent of the disk mass. As
a result, disks with lower initial masses will have lower crossover
radii.  Second, the settling time scale depends on the fluffiness of
the coagulating particles and this fluffiness remains uncertain. More
porous grains will settle more slowly and the cutoff radius $r_c$ will
decrease accordingly. As a result, we cannot completely rule out the
idea that photoevaporation of the solar nebula caused the observed
sharp cutoff in our Kuiper belt at 50 AU (if the Sun formed in the
high radiation environment of a large cluster).

\section{SUPPRESSION OF PLANET FORMATION IN CLUSTERS}  

The photoevaporation mechanism explored in this paper not only affects
the (possible) loss of gas from our own solar nebula, but also implies
that planet formation can be suppressed in other systems. In
particular, circumstellar disks associated with solar type stars can
be readily evaporated in sufficiently large clusters, whereas disks
around smaller (M type) stars can be evaporated in more common,
smaller groups.

As an observational example of a `large' cluster, consider the Hyades,
a relatively nearby stellar aggregate that is being searched for
planets. At the present time, the cluster mass is estimated to lie in
the range $M_C$ = 300 -- 460 $M_\odot$ (e.g., Perryman et al. 1998). 
The metallicity is relatively high, [Fe/H] = 0.14 $\pm$ 0.05. Given
the apparent correlation of extra-solar giant planets with metallicity
(e.g., Gonzales et al. 2001), we would expect the Hyades stars to
readily form giant planets in the absence of any disruption effects
from the background cluster.

To assess the effects of the cluster environment on planet formation
in the Hyades, we need to estimate its properties during its first 10
Myr of evolution when planets are expected to form. The Hyades now has
an age of $\sim 700$ Myr, so we must extrapolate back to its youth.
This transformation has been done (Kroupa 1995) and indicates that the
cluster had a mass $M \sim 1300 M_\odot$ and $\nstar \sim 3000$ at its
dynamical beginning. These cluster properties are roughly comparable
to those of the Trapezium cluster today (Hillenbrand \& Hartmann
1999), where circumstellar disks are observed to be actively
evaporating (see McCaughrean \& O'Dell 1996 and many others). In
addition, we note that $\nstar \sim 3000$ is the number of stars left
in the system after gas removal from the young cluster. During its
first few Myr of life, the cluster must contain even more stars $N_0 =
N_\star / {\cal F}_\star (\epsilon)$.  The fraction ${\cal F}_\star$
of stars remaining after gas removal depends on the star formation
efficiency of the cluster, but is expected to be ${\cal F}_\star \sim
3/4$ (see Adams 2000 for further detail). The formative stage of the
Hyades could thus have nearly $\nstar \sim 4000$ stellar members. In
estimating the radiation field of the cluster during its first 10 Myr
of life, we need to find the expected number of O and B stars in a
randomly selected population of $\nstar \sim 4000$ (where we assume
that the massive stars tend to form in the cluster center, as
observed, so that they are not likely to leave the cluster during the
gas removal adjustment phase). For a standard stellar IMF, a
collection of 4000 stars should contain $\sim$12 -- 16 stars with
$M_\ast > 8 M_\odot$ (large enough to explode as supernovae) and
should produce enough ultraviolet radiation to effectively evaporate
circumstellar disks on a short time scale.

Such large clusters can readily produce FUV radiation fields with
$G_0$ = 30,000, strong enough to affect planet formation. As shown in
Figure 9 for solar type stars, the disk evaporation time scale is
comparable to the expected planet formation time scale (10 Myr) for
disk radii $r_d \approx$ 15 AU (note that the disk accretion time
scale is also $\sim$10 Myr for $r_d$ = 15 AU and $\alpha = 10^{-4}$;
see Figure 8 and equation [\ref{eq:diffuse}]). Giant planet formation
could thus be inhibited in large clusters like the Hyades. In
addition, even though such planets could still form in the disk region
$r = 5-15$ AU, little disk mass (outside that region) would be
available to drive planet migration. Thus, FUV radiation fields can
alter the expected numbers and locations of planets for solar systems
forming in large clusters, and may account for the observed
underabundance (so far) of giant planets orbiting close to their
central stars in the Hyades (e.g., Paulson et al. 2004, Cochran,
Hatzes, \& Paulson 2003, Hatzes \& Cochran 2000). A similar deficit
of planets has been found in the globular cluster 47 Tucanae
(Gilliland et al. 2000), while populous metal-rich open clusters such
as NGC 6791 are currently being surveyed (Mochejska et al. 2002). 
We note that if giant planets form rapidly through gravitational
instability (e.g., Boss 2000), then we would not expect an
anti-correlation of giant planets with the strength of the radiation
fields.

The effect of FUV radiation on planet formation is more dramatic for
stars of lower mass. Figure 10 shows the evaporation time scale for
circumstellar disks exposed to a moderate FUV radiation field with
$G_0$ = 3000. The evaporation time decreases rapidly with the mass of
the central star, due to its (weaker) gravitational binding energy.
Stars with $M_\ast$ = 0.25 $M_\odot$, which lie near the peak of the
stellar mass distribution, will evaporate down to disk radii of 7 AU
during the 10 Myr time interval of planet formation. We thus
anticipate that giant planet formation can be seriously inhibited
around low mass stars. However, these considerations do not preclude
the formation of rocky terrestrial planets. The discussion of the
previous section indicates that dust grains can easily coagulate on
sufficiently short time scales to avoid being removed via
photoevaporation. If stars form in reasonably large ensembles, then
photoevaporation should remove gas, but not rocky dust grains, from
the mass reservoir available to form planets. This effect is strongest
for the least massive stars, so a clean prediction emerges: The
metallicity of planets should increase with decreasing mass of the
parental stars (for a given stellar metallicity). The magnitude of
this trend depends on the typical intensity of radiation fields in
star forming regions, and these radiation fields, in turn, depend on
the size and density of those regions.

\section{SUMMARY AND DISCUSSION}

In this paper, we have studied the photoevaporation of small
circumstellar disks ($r_d < r_g \sim 100$ AU) due to the heating by
FUV radiation from the stellar birth environment. Because this work
applies to small disk radii, we can determine the effects of
photoevaporation on inhibiting planet or planetesimal formation in the
disk region where $r = 10 - 100$ AU.  This work complements previous
studies, which have considered the evaporation of circumstellar disks
due to EUV radiation from their parental stars (e.g., Shu et al. 1993)
and the evaporation of large disks ($r_d \gta 100$ AU) due to UV
radiation in large clusters like the Trapezium (e.g., SH99).  We show
that FUV photoevaporation is likely to dominate EUV evaporation both
in large clusters (e.g., $N_\star \approx 4000$, $G_0 \approx 30,000$)
and in more moderate sized groups (e.g., $N_\star \approx 300$, 
$G_0 \approx 3000$), until the disks shrink to sizes $r_d \lta 10$
AU. By the time disks evaporate to such small radii (on time scales 
$t \gta 30$ Myr), the major episodes of planet formation are expected 
to be over, so that EUV photoevaporation does not generally play an
important role in affecting planet formation.

[1] For solar type stars, with $M_\ast \approx 1 M_\odot$, relatively
intense FUV radiation fields are required for significant
photoevaporation to take place. In particular, FUV radiation with
$G_0$ = 30,000 will efficiently evaporate disks with radii down to
$r_d \sim 20$ AU on time scales of $\sim10$ Myr.  The outer parts of
these circumstellar disks can be effectively evaporated through the
action of this level of FUV radiation, which is expected to be present
in the cores of dense stellar clusters (e.g., $d \lta 0.7$ pc with 
$N_\star \approx 4000$). 

[2] In our own solar system, the relative paucity of gas in Neptune
and Uranus can be understood if the outer solar nebula ($r \gta 20$
AU) is stripped of its gas before the planets complete their formation
(\S 5). The action of FUV radiation can remove enough gas on a
sufficiently rapid time scale if the early solar system is exposed to
FUV radiation fields with intensity $G_0 \ge 30,000$. We expect such
strong FUV radiation fields to be somewhat rare.

[3] FUV radiation fields can affect the formation of Kuiper belt
objects and other rocky bodies in the outer portion of our solar
system, and others. In these systems, dust grains coagulate as they
settle and eventually grow too large to be removed from the
disks. This process competes against evaporation, which acts to
remove gas and dust from the disk. We find that dust coagulation tends
to take place more rapidly (than mass loss) for radii less than a
cutoff radius $r_c \approx 100$ AU, even in relatively harsh stellar
birth environments (\S 6). As a result, Kuiper belt objects, and the
debris dust that they generate later on, can be formed (out to
$r\sim100$ AU) around most stars. However, we cannot completely rule
out the possibility that photoevaporation in the solar nebula could
have produced the observed cutoff in Kuiper Belt objects at $r_c \sim
50$ AU.

[4] Relatively large clusters contain B stars (and even O stars) with
high probability. Sufficiently rich clusters thus provide a hostile
environment for giant planet formation because the FUV radiation from
the background cluster is effective at removing gas from nebular
disks.  Applying this result to known clusters, such as the Hyades 
(\S 7), we find that giant planet formation can be compromised in 
such environments. 

[5] We have calculated (numerically) mass loss rates $\dot M$ as a
function of stellar mass $M_\ast$, disk radius $r_d$, and FUV
radiation field $G_0$. We also provide a simple analytic solution that
approximately shows the scaling of the mass loss rate with these
parameters. However, the analytic results are presented in terms of
the column density $N_C$ of the heated surface gas, which is assumed
to be isothermal with sound speed $a_s$. Comparison to PDR codes is
required to determine $N_C$ and $a_s$ for a given radiation field
$G_0$.

[6] The mass loss rate is significant for disk radii much smaller than
the critical radius, in particular for $r_d/r_g \gta 0.15$.  Previous
work assumed negligible mass loss for $r_d < r_g$, so this finding
increases the range of viable parameter space for mass loss. However,
the mass loss rate drops exponentially for $r_d \lta 0.15 r_g$,
scaling roughly as ${\dot M} \propto \exp[-r_g/2r_d]$.

[7] If a disk has enough viscosity, then viscous spreading of the
outer disk edge can affect photoevaporation. As a disk becomes smaller
in radius, its photoevaporation time increases whereas its viscous
spreading time decreases. As a result, disks will shrink down to the
size at which the two time scales are in balance (see Figure 8).  This
process tends to enhance the effectiveness of photoevaporation by
feeding new material into the outer disk where it can be efficiently
removed by the outflow. 

[8] Photoevaporation is most effective for disks surrounding stars of
low mass (\S 7). For example, a disk around an M dwarf with $M_\ast$ =
0.25 $M_\odot$ can be evaporated down to 10 AU in only 12 Myr when
exposed to a modest FUV radiation field with $G_0$ = 3000. Such
radiation intensities occur readily in moderately sized stellar
groups, those with $N_\star \sim 300$, which represent a common star
forming environment (e.g., Lada \& Lada 2003, Porras et al. 2003).

A intriguing result emerges from this consideration of disk
evaporation and the corresponding loss of planet forming potential for
stars with varying mass. High mass stars are efficient at evaporating
their own circumstellar disks and are thus not expected to harbor
planets. At the other end of the mass spectrum, red dwarfs easily lose
their disks due to photoevaporation in the presence of modest external
FUV radiation fields (e.g., $G_0$ = 3000), which are expected in
common star forming units. As a result, solar type stars (loosely
speaking, stars with masses within a factor of two of 1.0 $M_\odot$)
are the preferred locations for giant planet formation. 

\bigskip 
\centerline{\bf Acknowledgments} 
\medskip 

We would like to thank M. Kaufman and A. Parravano for useful
discussions. We also thank an anonymous referee for comments that
clarified the paper. This work was supported by a grant from the NASA
Origins of the Solar System Program, the NASA Astrophysics Theory
Program, and by the Michigan Center for Theoretical Physics.

\newpage
\centerline{\bf APPENDIX: ANALYTIC APPROXIMATION} 
\centerline{\bf FOR THE PHOTOEVAPORATION OF SMALL DISKS ($r_d \ll r_g$)} 
\bigskip 

In this Appendix, we derive simple analytic results for the scaling of
the photoevaporative mass loss rate $\dot M$ as a function of stellar
mass $M_\ast$, disk radius $r_d$, and (implicitly) the strength of the
FUV radiation field $G_0$. Specifically, we make the following
simplifying assumptions:

[i] The gas is essentially static in the inner region where $r < r_s$,
with thermal pressure balancing gravity. This assumption is equivalent
to neglecting the $v dv/dr$ term in equation (\ref{eq:forcezero}) and
solving the remaining equation for the density structure $n(r)$.

[ii] The outflow velocity $v$ is constant in the outer region where 
$r > r_s$, with $v=a_s$, the sound speed at the sonic point. This 
assumption implies that the density profile in the outer region has
the form $n(r) = n_s (r_s/r)^2$, where $n_s$ is the number density at
$r_s$.

[iii] The FUV field $G_0$ heats a column $N_C$ of surface gas to a
constant temperature $T_s$ (i.e., the surface layer is isothermal).
We thus obtain results that depend on $T_s$ ($a_s$) and $N_C$, but
these parameters are actually surrogates for the radiation field
$G_0$.  We can relate $a_s$ and $N_C$ to $G_0$ using the results 
from the PDR code as shown in Figure 2. 

Figure 1 shows a schematic representation of the photoevaporation
process for subcritical disks. Here, we work in the limit $r_d \ll
r_g$ and assume that the critical radius $r_g$ and the sonic radius 
$r_s$ are comparable ($r_s \sim r_g$). With this set of approximations, 
the density profile for the subsonic region takes the form 
$$
n(r) = n_d \exp\Bigl[ - {r_g \over 2 r_d} (1 - r_d/r)^2 \Bigr] 
\, , \eqno({\rm A}1) 
$$
where $r_g$ is the critical radius (given by eq. [\ref{eq:rcrit}]) 
with $T=T_s$. Assuming that $r_s, r_g \gg r_d$, we thus obtain 
$$
n_s \approx n_d \exp[-r_g/2r_d] \, . \eqno({\rm A}2) 
$$
The mass loss rate from the disk edge (at $r_d$) is given by the 
continuity equation and takes the form 
$$
{\dot M} = \muc n_s a_s {\cal A}_s \, , \eqno({\rm A}3) 
$$
where $\muc$ is the mass per particle and where ${\cal A}_s$ is 
the area subtended by the flow at $r_s$. This area can be written 
$$
{\cal A}_s = 2 \pi r_d H_d (r_s/r_d)^2 \equiv 
2 \pi \alpha r_g (r_d r_g)^{1/2} \, . \eqno({\rm A}4) 
$$ 
In the second equality, we have evaluated the disk scale height 
$H_d$ = $r_g (r_d/r_g)^{3/2}$ and have defined a dimensionless 
constant $\alpha \equiv (r_s/r_g)^2$, which is of order unity. 
Finally, we apply the condition that the external FUV flux $G_0$ 
heats a column density $N_C$ given by the integral 
$$
N_C = \int_{r_d}^\infty n(r) dr \, . \eqno({\rm A}5) 
$$ 
In the limit that $r_d \ll r_g$, most of the support of this 
integral occurs for small $r$ where equation (A1) applies. 
This condition (A5) relates the column density $N_C$ to $n_d$, 
and, to leading order, this relation takes the form
$$
n_d \approx \Bigl( {2 \over \pi} \Bigr)^{1/2}  
\Bigl( {r_g \over r_d} \Bigr)^{1/2} 
{N_C \over r_d} \, . \eqno({\rm A}6) 
$$ 
Collecting all of the results given above, we obtain the 
following expression for the mass loss rate 
$$
{\dot M} = C_0 N_C \muc a_s r_g \Bigl( {r_g \over r_d} \Bigr)
{\rm e}^{-r_g/2r_d} \, , \eqno({\rm A}7)  
$$ 
where $C_0$ is a dimensionless constant of order unity. 

Although the derivation of equation (A7) applies only in the limit
$r_d \ll r_s \sim r_g$, the resulting function can be evaluated when
$r_d \approx r_s \approx r_g$ and implies nearly the same result as
the supercritical mass loss rate of \S 3 (see eq. [\ref{eq:mdotzero}]).
Therefore, we can use equation (A7) as an analytic approximation to
the mass loss rate (for a given radiation field $G_0$) as a function
of $r_d/r_g$ (for $r_d/r_g \le 1$).  This approximation should match
onto the subcritical mass loss rates calculated in \S 4 (where
$r_d/r_g \approx 0.125$) and should also match onto the supercritical
mass loss rates of \S 3 (for $r_d \to r_g$). Notice that we are
implicitly assuming that $a_s$ and $N_C$ do not change with $r_d/r_g$
for a given radiation field.

With these approximations, we can estimate the mass loss rates for
systems that are intermediate between the subcritical regime of \S 4
and the supercritical regime of \S 3. We can also use the resulting
form of $\dot M$ to understand how the mass loss rate depends on the
various parameters in the problem. As $r_d$ becomes comparable to 
$r_g$, the mass loss rate approach its supercritical value. As 
$r_d/r_g$ decreases, the mass loss rate decreases, but only slowly 
at first. The outflow rate $\dot M$ is half its supercritical value 
when $r_d/r_g \approx 0.17$ (significantly below unity). For even 
smaller values of $r_d/r_g$, however, the decaying exponential 
behavior wins and the mass loss rates drop dramatically. 

Finally, we can also make an analytic estimate for the mass loss rate
from the disk surface, i.e., for vertical flow off the top and bottom
of the disk. This estimate can be compared to that for mass loss from
the disk edges (see eq. [A7]). For vertical flow, we treat each
increment of disk surface area $2 \pi r dr$ with the same formulation
used above for the disk edges, with one exception: We must replace the
radius $r_d$ with $r \le r_d$ and then integrate over $r$.  This
procedure takes into account the fact that material at $r < r_d$ lives
deeper in the gravitational potential well and is harder to extract 
from the system. The resulting mass loss rate from the disk surface is 
$$
{\dot M}_{sur} = C_1 N_C \muc a_s r_g 
\Bigl( {r_g \over r_d} \Bigr)^{1/2} {\rm e}^{-r_g/2r_d} 
\, , \eqno({\rm A}8) 
$$ 
where all of the dimensionless quantities are collected into the
constant $C_1$ (which is comparable to, but not quite the same as, the
constant $C_0$ appearing in eq. [A7]).  Comparing the mass loss rates
from the disk edge and the disk surface, we find that ${\dot 
M}_{sur}/{\dot M} \approx (r_d/r_g)^{1/2}$. In the limit $r_d/r_g \ll
1$, the mass loss rate from the disk edge dominates the mass loss rate
from the disk surfaces.

\newpage 
\begin{figure}
\figurenum{1}
\epsscale{1.0}
\plotone{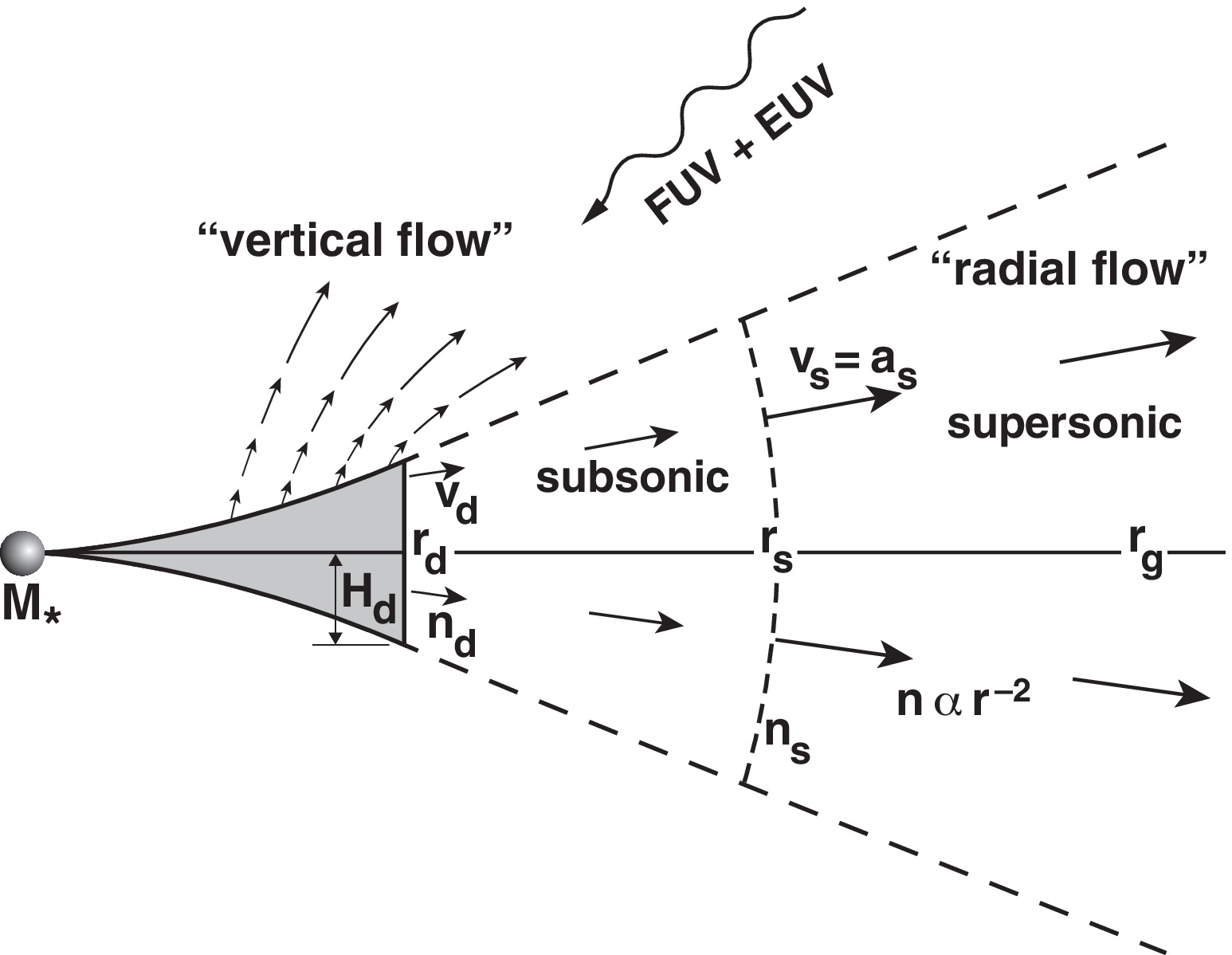}  
\figcaption{Schematic of a disk with radius $r_d$ around a star with 
mass $M_\ast$, illuminated by the FUV (and perhaps EUV) radiation from
nearby stars of greater mass. The disk is inclined so that the top and
edge are exposed. The disk scale height is $H_d$ at the outer radius
$r_d$. In the subcritical regime, where $r_d < r_g$, the bulk of the
photoevaporation flow (the radial flow) originates from the disk edge,
which marks the inner boundary. The flow begins subsonically at $r_d$,
with speed $v_d$ and density $n_d$. The flow accelerates to the sound
speed at $r_s$ (the sonic point), which lies inside the critical
escape radius $r_g$. Beyond the sonic point, the flow attains a
terminal speed of order the sound speed and the density falls roughly
as $n \propto r^{-2}$. Although some material is lost off the top and
bottom faces of the disk (the vertical flow), its contribution to the
mass loss rate is secondary to that from the edges. Nonetheless, the
polar regions are not evacuated, the star is fully enveloped by
circumstellar material, and the incoming FUV radiation will be
attenuated in all directions. }
\end{figure} 

\newpage 
\begin{figure} 
\figurenum{2a}
\epsscale{1.0}
\plotone{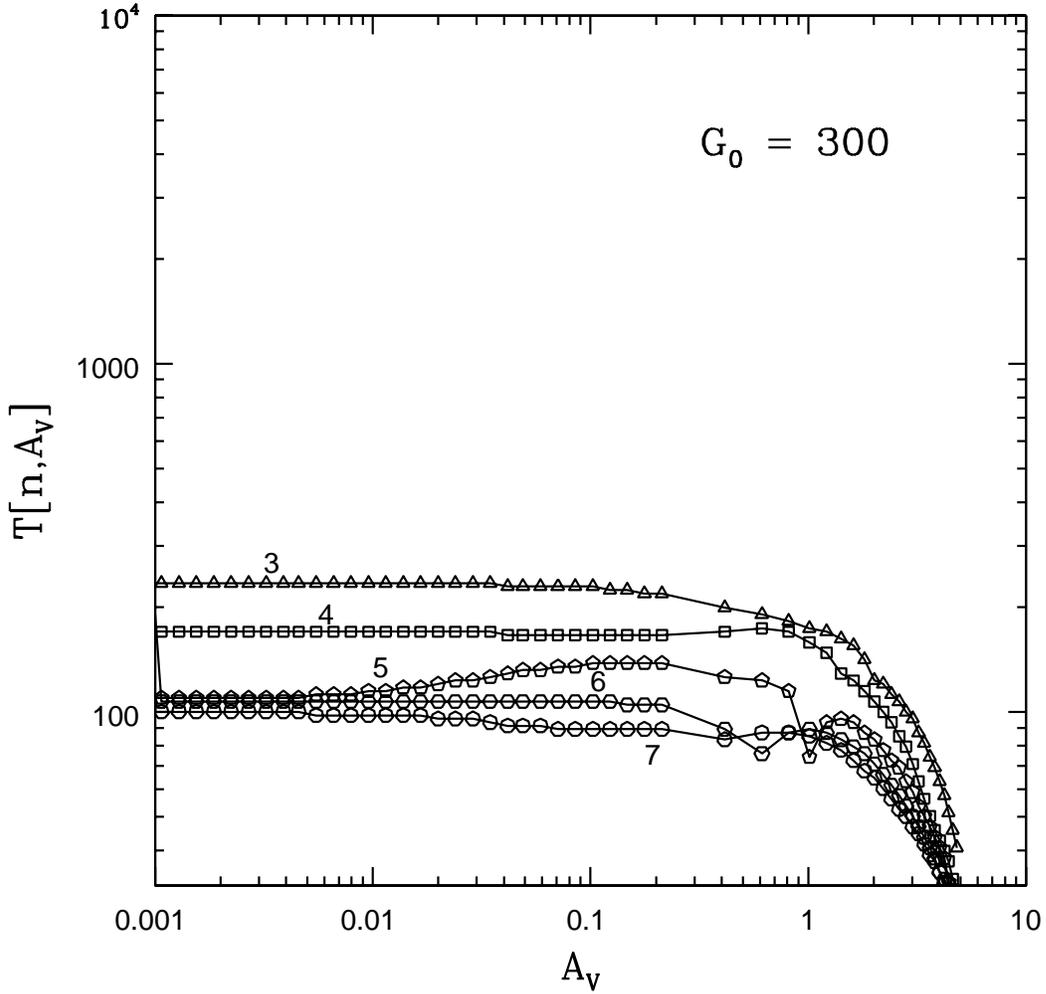}
\figcaption{Temperature profiles calculated from the PDR code 
for an external FUV radiation field. Each curve shows 
the temperature as a function of visual extinction $A_V$ for a given  
number density $n = 10^p$ cm$^{-3}$, shown here for $p = 3 - 8$. 
The values of $p$ are labeled for each curve; in addition, each 
curve is marked by polygons, where the number of sides corresponds to 
the value of $p$. (a) Results for $G_0$ = 300. (b) Results for 
$G_0$ = 3000. (c) Results for $G_0$ = 30,000. }
\end{figure} 

\newpage 
\begin{figure} 
\figurenum{2b}
\epsscale{1.0}
\plotone{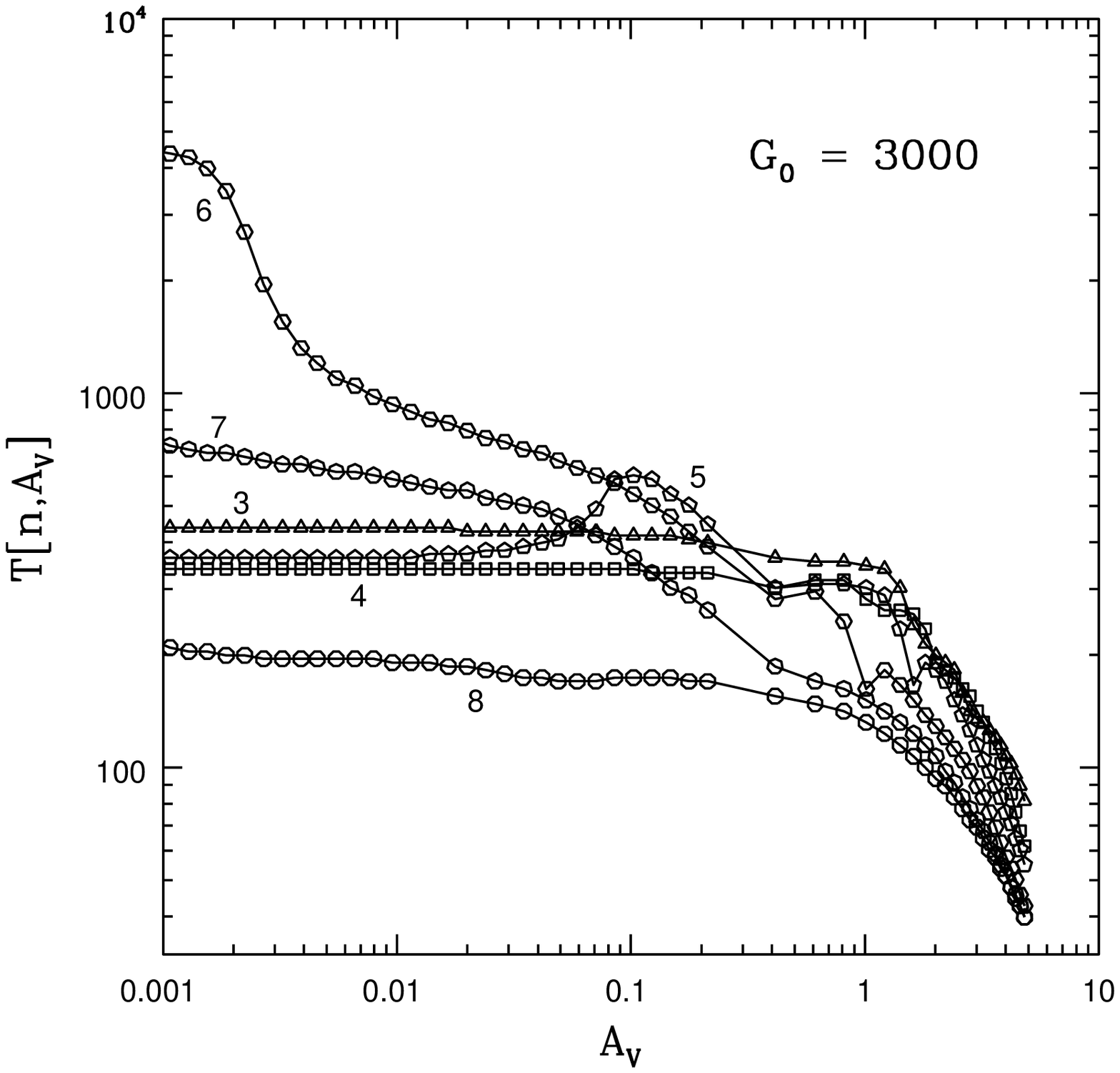}
%\figcaption{Temperature profiles calculated from the PDR code 
%for an external radiation field with $G_0$ = 3000. Each curve shows 
%the temperature as a function of visual extinction $A_V$ for a given 
%number density $n = 10^p$ cm$^{-3}$, shown here for $p$ = 3, 4, 5, 6,
%7, and 8. The values of $p$ are labeled for each curve; in addition, 
%each curve is marked by polygons, where the number of sides
%corresponds to the value of $p$. }
\end{figure} 

\newpage 
\begin{figure} 
\figurenum{2c}
\epsscale{1.0}
\plotone{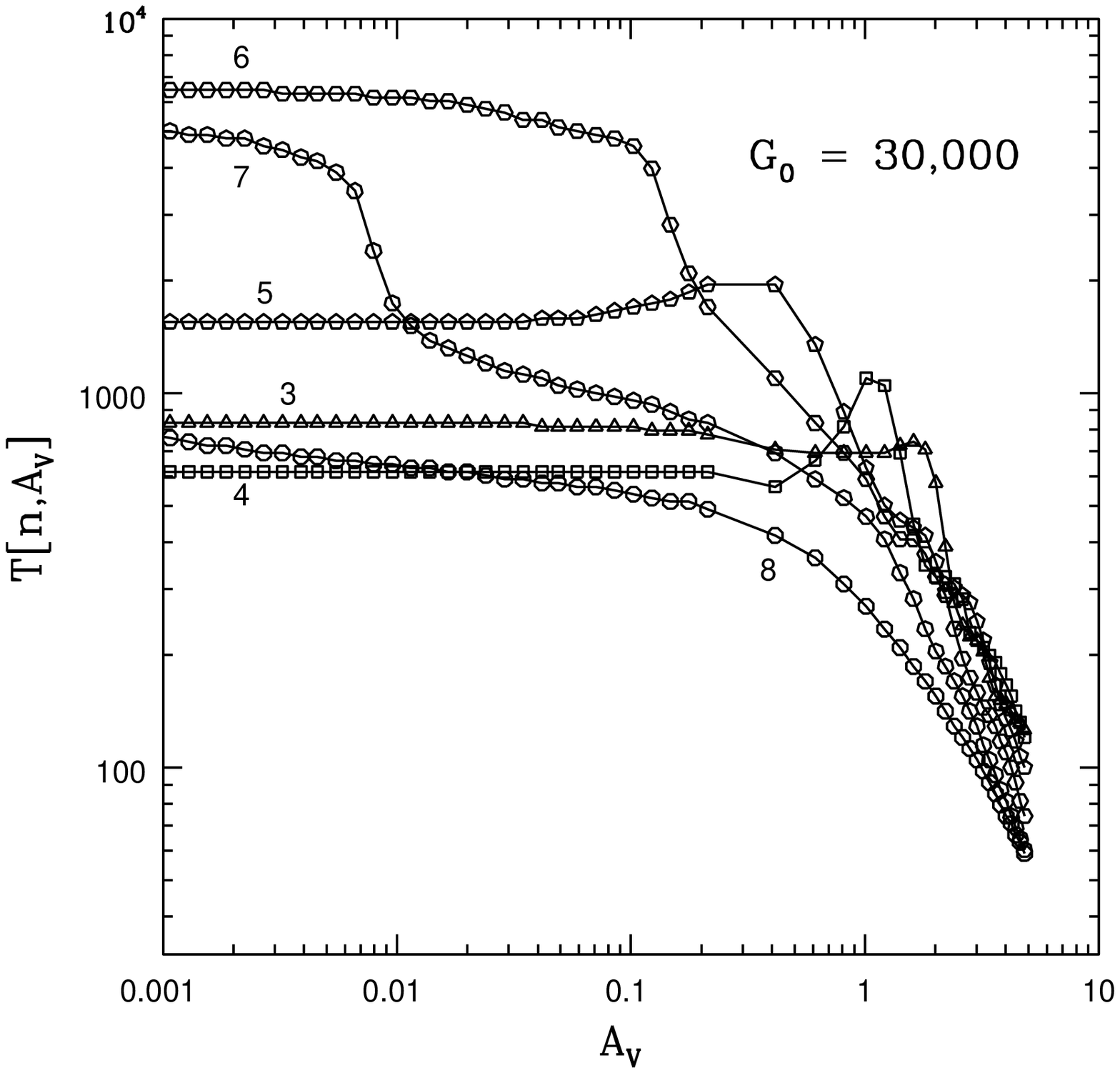}
%\figcaption{Temperature profiles calculated from the PDR code 
%for an external radiation field with $G_0$ = 30,000. Each curve shows 
%the temperature as a function of visual extinction $A_V$ for a given 
%number density $n = 10^p$ cm$^{-3}$, shown here for $p$ = 3, 4, 5, 6,
%7, and 8. The values of $p$ are labeled for each curve; in addition, 
%each curve is marked by polygons, where the number of sides
%corresponds to the value of $p$. } 
\end{figure} 

\newpage 
\begin{figure} 
\figurenum{3}
\epsscale{1.0}
\plotone{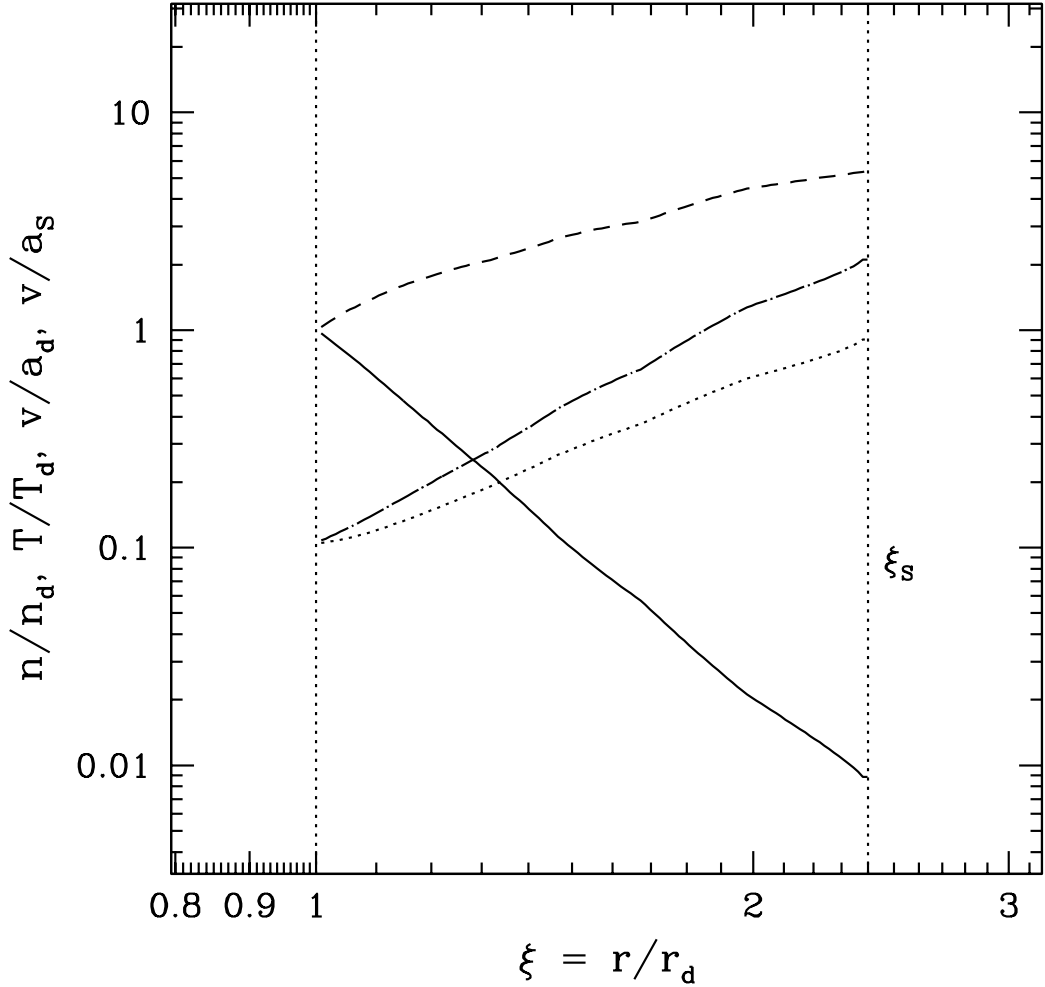}
\figcaption{Radial profiles of the fluid fields for a disk with 
$r_d$ = 30 AU surrounding a solar mass star ($M_\ast$ = 1.0
$M_\odot$), where the radiation field has intensity $G_0$ = 3000.  
The solid curve shows the run of density; the dashed curve shows the
temperature; the dot-dashed curve shows the outflow speed. All of
these quantities are normalized to their values at the outer disk
edge.  Also shown, depicted as the dotted curve, is the Mach number
$\cal M$ = $v/a_S$, which reaches unity at the sonic point (by
definition). Finally, the vertical lines at $\xi = 1, \xi_s$ mark 
the locations of the disk edge and the sonic point, respectively.} 
\end{figure} 

\newpage 
\begin{figure}
\figurenum{4}
\epsscale{1.0}
\plotone{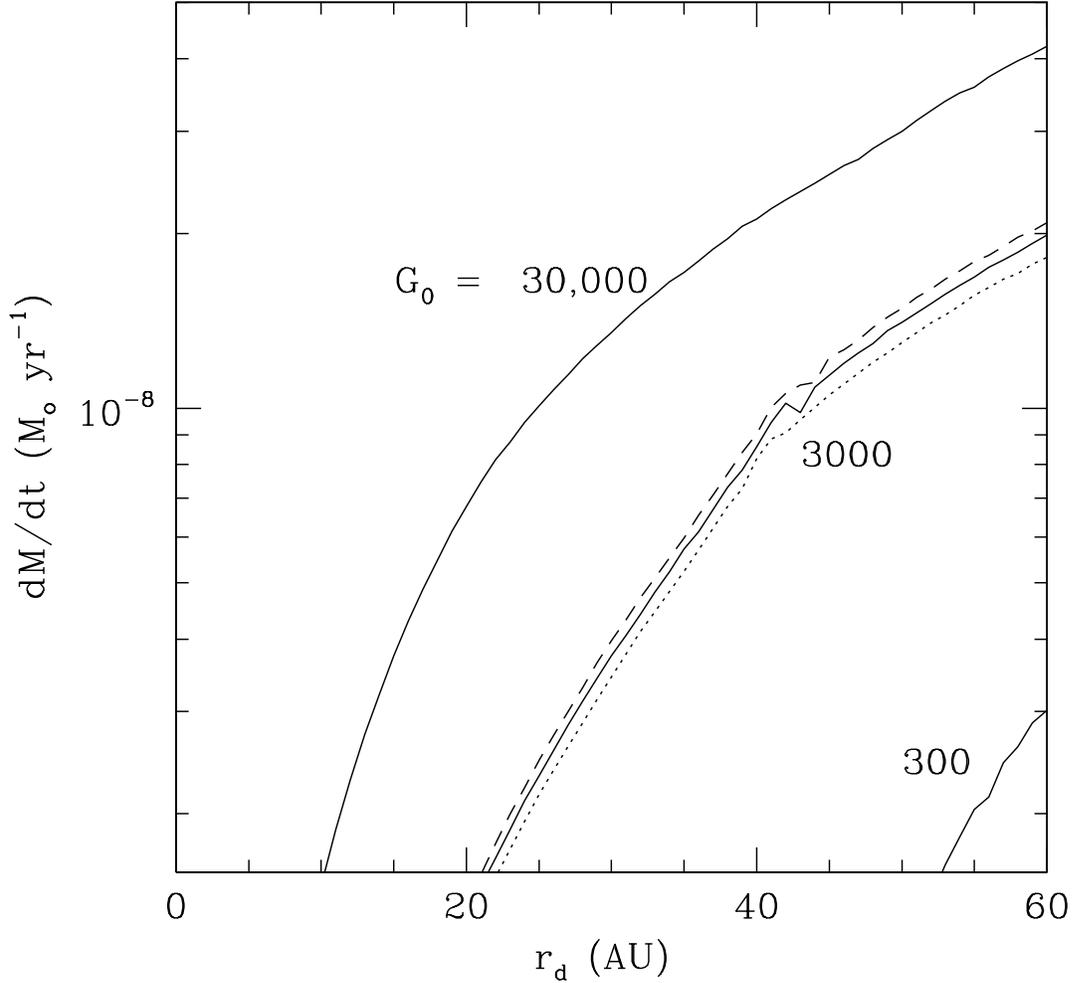} 
\figcaption{The mass loss rates due to photoevaporation for a 
circumstellar disk embedded in FUV radiation fields of varying
intensities (as labeled). The mass loss rates are shown as a function
of disk radius $r_d$ for a fixed stellar mass $M_\ast$ = 1.0 $M_\odot$. 
For the radiation field with intensity $G_0$ = 3000, we show the
effects of varying the inner boundary condition. The solid curve uses
our standard choice of inner boundary condition where $T_d$ = 75 K at
$r_d$ = 30 AU; the dashed curve shows the alternate choice of $T_d$
(30 AU) = 90 K; the dotted curve uses $T_d$ (30 AU) = 60 K. In all 
cases, the disk ``surface'' temperature (in the absence of external 
flux) is assumed to follow a power-law of the form 
$T_d = T_d$ (30 AU) (30 AU/$r)^{1/2}$. } 
\end{figure}  

\newpage 
\begin{figure}
\figurenum{5}
\epsscale{1.0}
\plotone{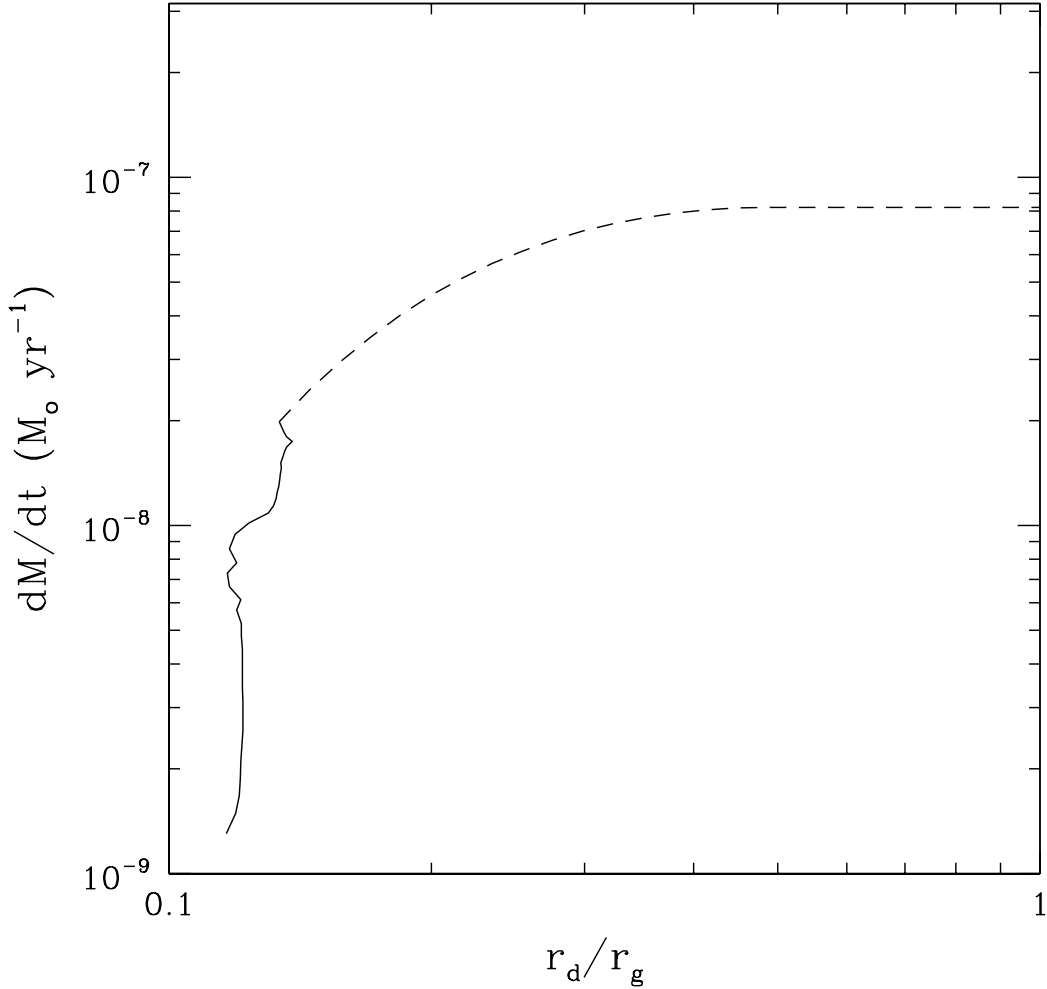}
\figcaption{The mass loss rate (in $M_\odot$ yr$^{-1}$) is shown as a 
function of $r_d/r_g$ for a disk surrounding a $M_\ast = 1.0 M_\odot$
star with an external FUV radiation $G_0$ = 3000. The solid portion of
the curve shows results from our numerical treatment, which spans a
wide range in $\dot M$, but a relatively narrow range in $r_d/r_g$. 
The dashed portion of the curve shows the analytic estimate (see the
Appendix), which smoothly joins that numerical result (for subcritical
disks) onto the result for supercritical disks. The numerical results
do not follow a smooth curve because the temperature $T_s$ at the
sonic point (and hence $r_g$) depends on $r_d$ through the complicated
PDR dependence of $T_s$ on $n$ and $N_H$ (see Fig. 2). Here we assume
a constant $r_g$ for the analytic (dashed) portion of the curve. }
\end{figure} 

\newpage 
\begin{figure}
\figurenum{6}
\epsscale{1.0}
\plotone{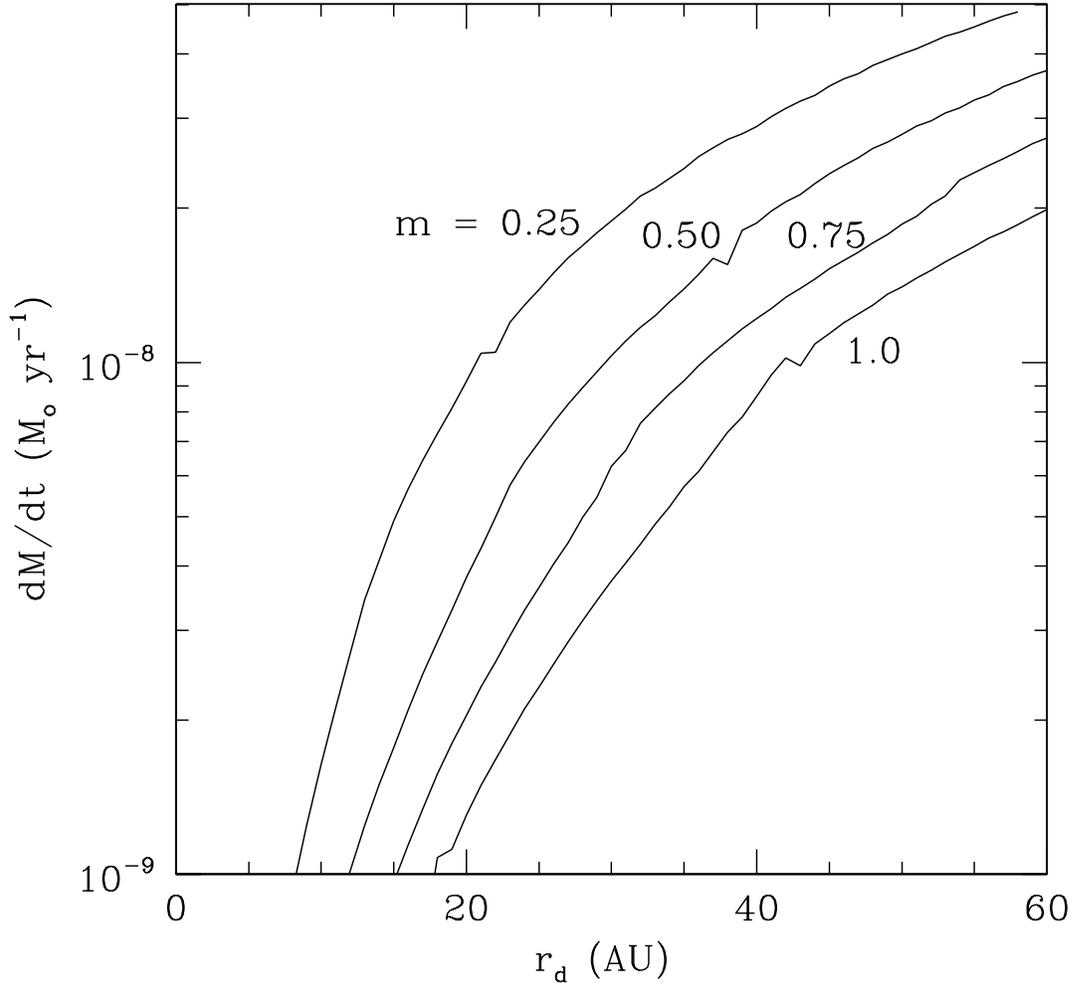}
\figcaption{The mass loss rates due to photoevaporation for 
circumstellar disks embedded in an FUV radiation field with $G_0$ =
3000.  The mass loss rates are shown as a function of disk radius
$r_d$ for varying masses of the central stars, where $m = M_\ast/
M_\odot$ = 0.25, 0.50, 0.75 and 1.0 (as labeled).} 
\end{figure} 

\newpage 
\begin{figure}
\figurenum{7}
\epsscale{1.0}
\plotone{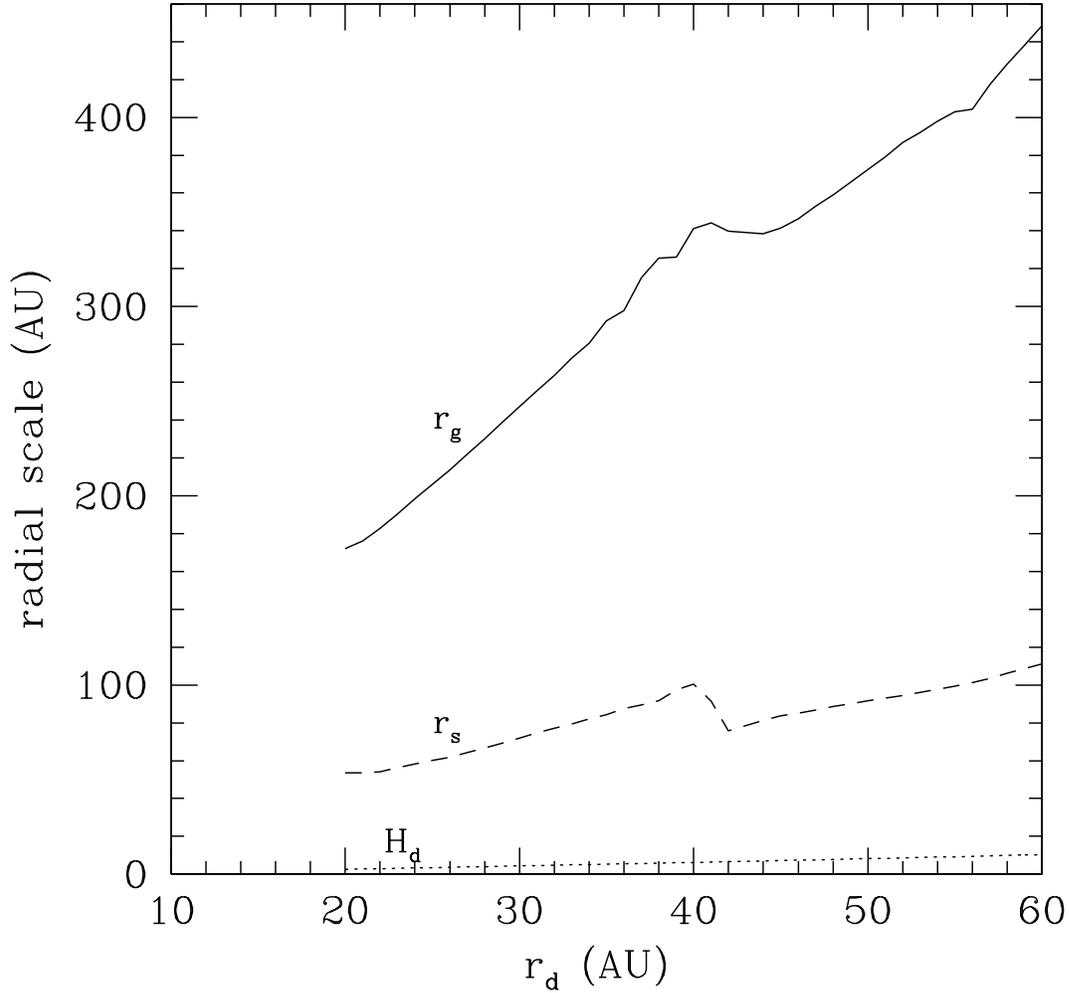}
\figcaption{This figure shows the relevant length scales as 
a function of disk radius $r_d$ for a circumstellar disk in orbit
about a star with mass $M_\ast$ = 1.0 $M_\odot$ and exposed to an FUV
radiation field of intensity $G_0$ = 3000. The dotted curve shows the
disk scale height $H_d$. The dashed curve shows the sonic radius $r_s$
as a function of disk radius. The solid curve shows the critical
radius $r_g$. The structure in the latter two curves is a reflection
of the structure in the relationship between temperature and visual
extinction (see Fig. 2). }
\end{figure} 

\newpage 
\begin{figure}
\figurenum{8}
\epsscale{1.0}
\plotone{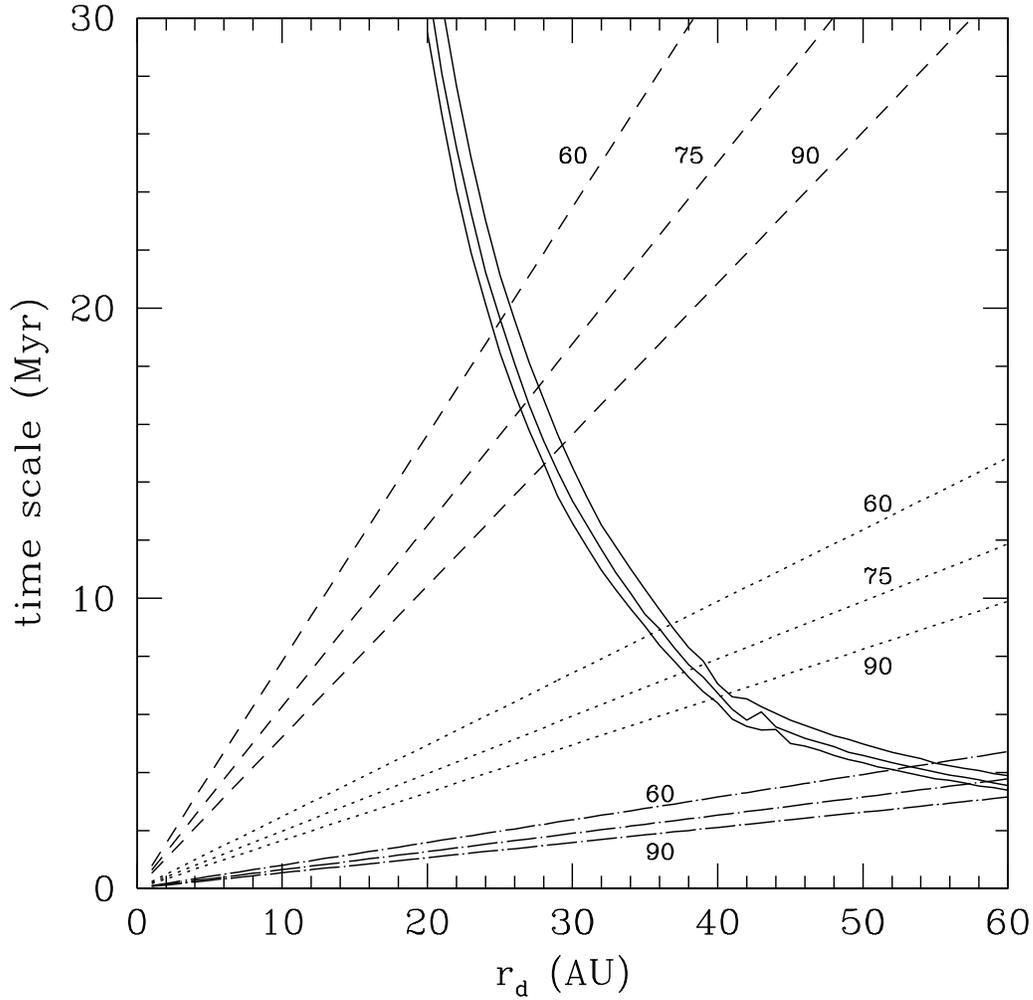}
\figcaption{Comparison of evaporation time scales and disk accretion 
time scales for a disk surrounding a star with mass $M_\ast$ = 1.0
$M_\odot$. The FUV radiation field has $G_0$ = 3000.  The solid curve
shows the evaporation time scale, as calculated in this paper, which
is a decreasing function of disk radius. The disk accretion time is an
increasing function of disk radius. Results are shown for viscosity
parameter $\alpha = 10^{-4}$ (dashed lines), $\alpha=10^{-3.5}$
(dotted lines), and $\alpha = 10^{-3}$ (dot-dashed lines). Each case
has three separate curves which correspond to different choices for
the disk temperature scale, i.e., the temperature $T_d$ ($r_d$ = 30
AU); the labels refer to $T_d$(30 AU) in Kelvin. For a given value of
the viscosity parameter $\alpha$, the disk will shrink down to the
radius where the evaporation time equals the accretion time. }
\end{figure} 

\newpage 
\begin{figure}
\figurenum{9}
\epsscale{1.0}
\plotone{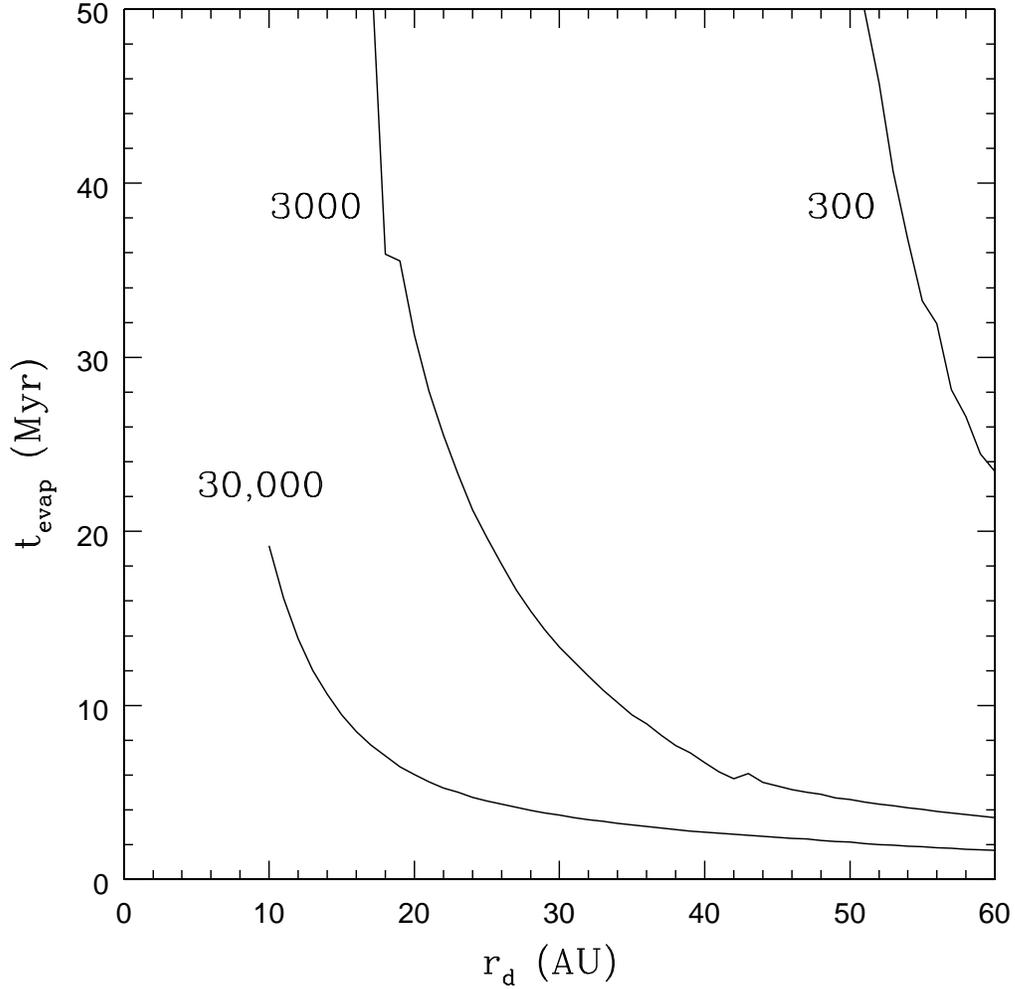}
\figcaption{Photoevaporation time scales for circumstellar disks 
exposed to varying external radiation fields, $G_0$ = 300, 3000, and
30,000 (as labeled). These models assume that the disk orbit around
central stars with mass $M_\ast$ = 1.0 $M_\odot$. The evaporation
rates are calculated according to the formulation of this paper. 
To specify the evaporation time scale, we assume that disks have 
masses $M_d$ = 0.05 $M_\ast$ ($r_d$/30 AU)$^{1/2}$. The evaporation 
time scales, as plotted, are proportional to the assumed disk mass, 
$\tevap \propto M_d$. } 
\end{figure} 

\newpage 
\begin{figure}
\figurenum{10}
\epsscale{1.0}
\plotone{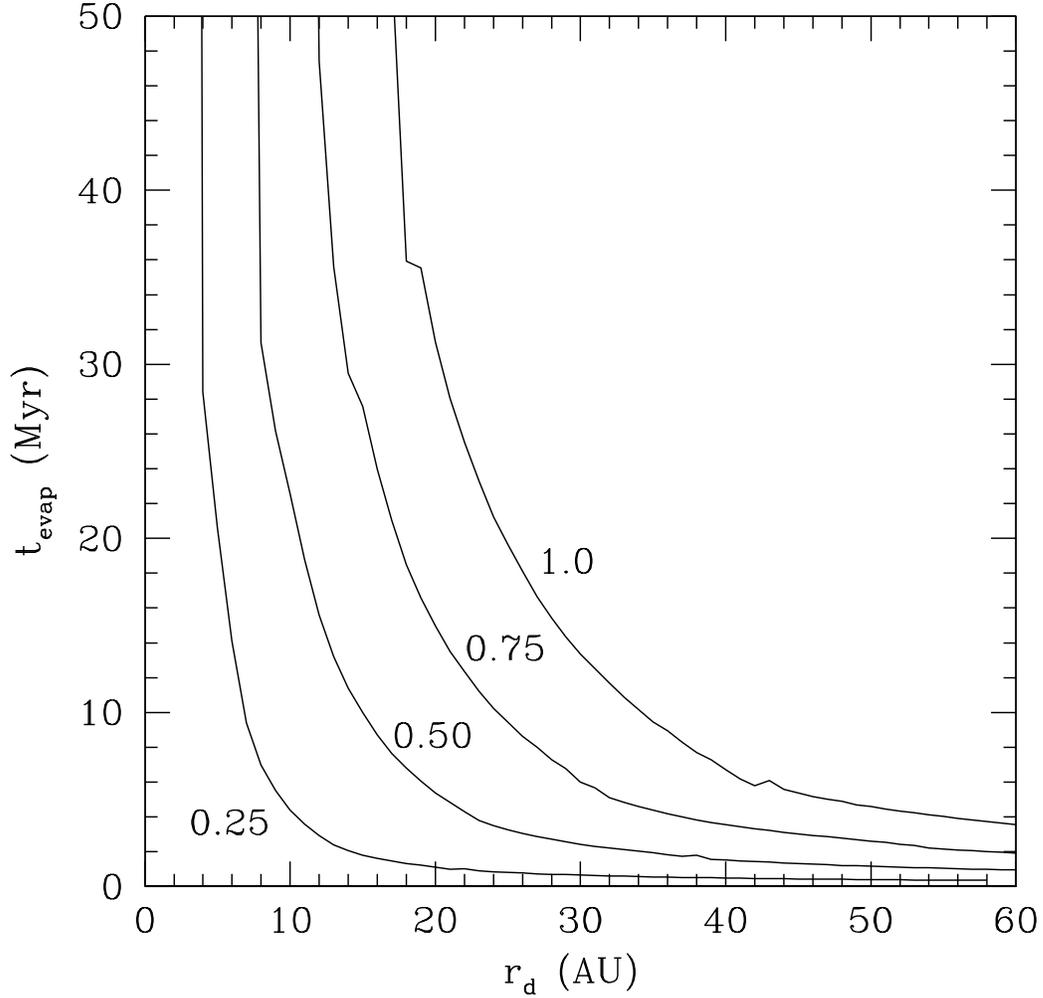}
\figcaption{Photoevaporation time scales for circumstellar disks 
surrounding stars with varying masses $m = M_\ast/ M_\odot$ = 1.0,
0.75, 0.5, and 0.25 (as labeled). The external FUV radiation field has
intensity $G_0$ = 3000 for all cases shown.  The evaporation rates are
calculated according to the formulation of this paper. To specify the
evaporation time scale, we assume that disks have masses $M_d$ = 0.05
$M_\ast$ ($r_d$/30 AU)$^{1/2}$. The evaporation time scales, as plotted, 
are proportional to the assumed disk mass, $\tevap \propto M_d$. }
\end{figure}

\end{document}